\documentclass[showpacs,twocolumn,pra,superscriptaddress,notitlepage]{revtex4-1}
\usepackage{qcircuit}
\usepackage[dvips]{graphicx}
\usepackage{amsmath,amssymb,amsthm,mathrsfs,amsfonts,dsfont}
\usepackage{subfigure, epsfig}
\usepackage{braket}
\usepackage{bm}
\usepackage{fancyhdr}
\usepackage{enumerate}
\usepackage{color}
\usepackage{physics}
\usepackage{pdfpages}
\usepackage{pgffor}

\newtheorem{theorem}{Theorem}

\newtheorem{proposition}{Proposition}

\usepackage[dvipsnames]{xcolor}
\newcommand{\comments}[1]{}
\newcommand{\sun}{\textcolor{black}}

\newcommand{\Han}[1]{\textcolor{black}{#1}} 
\newcommand{\Zimu}[1]{\textcolor{black}{#1}}

\color[rgb]{0.2,0.2,0.2} %grey
\makeatletter
\AtBeginDocument{\let\LS@rot\@undefined}
\makeatother
\begin{document}
% \title{Towards a variational Jordan-Lee-Preskill quantum algorithm}
\title{Towards a variational Jordan-Lee-Preskill quantum algorithm}

\begin{abstract}
Rapid developments of quantum information technology show promising opportunities for simulating quantum field theory in near-term quantum devices. In this work, we formulate the theory of (time-dependent) variational quantum simulation of the 1+1 dimensional $\lambda \phi^4$ quantum field theory including encoding, state preparation, and time evolution, with several numerical simulation results. These algorithms could be understood as near-term variational quantum circuit (quantum neural network) analogs of the Jordan-Lee-Preskill algorithm, the basic algorithm for simulating quantum field theory using universal quantum devices. Besides, we highlight the advantages of encoding with harmonic oscillator basis based on the LSZ reduction formula and several computational efficiency such as when implementing a bosonic version of the unitary coupled cluster ansatz to prepare initial states. We also discuss how to circumvent the ``spectral crowding" problem in the quantum field theory simulation and appraise our algorithm by both state and subspace fidelities.

\end{abstract}
\date{\today}

 \author{Junyu Liu$^*$}
 \email{junyuliu@uchicago.edu}
 \affiliation{Walter Burke Institute for Theoretical Physics, California Institute of Technology, Pasadena, CA 91125, USA}
 \affiliation{Institute for Quantum Information and Matter, California Institute of Technology, Pasadena, CA 91125, USA}
 \affiliation{Pritzker School of Molecular Engineering, The University of Chicago, Chicago, IL 60637, USA}
 \affiliation{Chicago Quantum Exchange, Chicago, IL 60637, USA}
 \affiliation{Kadanoff Center for Theoretical Physics, The University of Chicago, Chicago, IL 60637, USA}

\author{Zimu Li}
\email{lizm@mail.sustech.edu.cn}
\affiliation{DAMTP, Center for Mathematical Sciences, University of Cambridge, Cambridge CB30WA, UK}

\author{Han Zheng}
\email{hanz98@uchicago.edu}
\affiliation{Department of Statistics, The University of Chicago, Chicago, IL 60637, USA}

\affiliation{DAMTP, Center for Mathematical Sciences, University of Cambridge, Cambridge CB30WA, UK}

 \author{Xiao Yuan}
 \affiliation{Center on Frontiers of Computing Studies, Peking University, Beijing 100871, China}
 \affiliation{Stanford Institute for Theoretical Physics, Stanford University, Stanford, CA 94306, USA}

 \author{Jinzhao Sun$^*$}
 \email{jinzhao.sun@physics.ox.ac.uk}
 \affiliation{Clarendon Laboratory, University of Oxford, Parks Road, Oxford OX1 3PU, United Kingdom}

\maketitle
*: corresponding authors.
\section{Introduction}
Quantum information science is currently an important direction of modern scientific research across several subjects, including quantum physics, computer science, information technology, and quantum engineering. The rapid development of quantum technology brings us evidence that quantum computers in the near-term are able to perform some specifically scientific computations using dozens of qubits, but errors appearing in the noisy quantum circuits might set certain limits of the computational scale \cite{preskill2018quantum,arute2019quantum}. At the current stage, it makes sense to assume a reasonable quantum device exists and study potential scientific applications of such a device. This forms one of the main topics in the modern research of quantum information science. 

Among numerous quantum applications, physicists, in particular, might care about how quantum devices could enlarge the range of computational capacity on certain problems in fundamental physics. In modern physics, quantum field theory is a general language or paradigm for describing almost all phenomena existing in the world, from sub-atomic particle physics, string theory and gravity, to condensed-matter and cold-atomic physics. If we could imagine the existence of powerful quantum computers, it will be natural, important, and interesting to consider if quantum computation could address open problems appearing in the study of quantum field theories, where many of them are at strong coupling and strong correlation. In fact, simulating quantum field theories in quantum devices is one of the earliest motivations of quantum computation \cite{feynman1982simulating}, and becomes an important new research direction recently in the physics community, see the references \cite{Preskill:2018fag,cyber,Jordan:2011ne,Jordan:2011ci} as examples. 

When simulating quantum field theories, or more generally, solving some well-defined computational tasks using quantum computation, theorists will either assume a universal, fault-tolerant quantum computer, or a noisy, near-term quantum circuit without enough quantum error correction. Both of them are wise choices and important scientific directions. Using fault-tolerant quantum computing is helpful for theoretical, conceptual problems or development of quantum devices usually appearing in the long-term, while studying near-term, early quantum computation will allow us to use existing machines and do experiments. In this paper, we will focus on the second direction, by exploring how far quantum simulation could go using near-term devices, with the help of specific problems in quantum field theories. It is helpful to see the usage and limitations of the currently existing, or future possible quantum hardware to simulate quantum field theories, and benchmark our quantum devices using interesting problems in fundamental physics \cite{Milsted:2020jmf}. Eventually, we believe that a universal, fault tolerant quantum device will come true, and we believe that our work might be helpful to speed up the process. 

Here, we are specifically looking at the Jordan-Lee-Preskill scattering problem \cite{Jordan:2011ne,Jordan:2011ci} in the 1+1 dimensional $\lambda \phi^4$ quantum field theory. The research about scattering problems has a long history in physics, from the scattering experiment of alpha particles by Rutherford to the modern discovery of the Higgs boson. Performing scattering experiments and determining scattering matrices are important themes in particle physics and quantum field theories. In Refs.~\cite{Jordan:2011ne,Jordan:2011ci}, Jordan, Lee, and Preskill designed a full algorithm running in a universal quantum computer to perform particle scattering in quantum field theories, containing initial state preparation, time evolution, and measurement, where the proof of polynomial complexity is presented. In this work, we will construct closely-related algorithms that are more suitable for near-term quantum computers. 

We will be most interested in the circumstance where we have a machine to perform variational quantum simulation and hybrid quantum-classical calculations (see, for instance, Refs.~\cite{Peruzzo2014,farhi2014quantum,McClean2016,li2017efficient,yuan2019theory,sugurureview21,cerezo_variational_2021,zhang2020low,Endo20Variational,XU2021}). In those algorithms, we will imagine that quantum gates or states are parametrized by a few parameters, and we iteratively perform measurements from quantum states and construct variational algorithms to optimize those parameters. We believe that those algorithms realized in the laboratory might be able to perform useful computations and could tell us something unknown about fundamental physics. In this work, we will systematically evaluate the possibility of variational quantum simulation in the context of $\lambda \phi^4$ quantum field theory. The paper is organized as following:

\begin{itemize}
	\item Basis choice. We will make a detailed comparisons between the field basis and the harmonic oscillator basis, momentum space, and coordinate space in Section~\ref{stateencoding}. All those choices have pros and cons. The field basis will cause the field correlations to be easy to measure and make the Lagrangian density local in coordinate space, but it will not be directly connected to the Feynmann rules and scattering calculations in the momentum space. Moreover, finding the eigenstates (for instance, the vacuum and low-lying one-particle states) might be not easy. It requires non-trivial digital quantum algorithms (with truncation error) for encoding. On the other hand, the harmonic oscillator basis is easy to formulate, track, and identify the energy levels of states, but may not be easy to identify field profiles. The Hamiltonian is non-local but still sparse in the momentum space. In this paper, we will work exclusively with harmonic oscillator basis (HO basis for brevity) as opposed to the field basis used in the original Jordan-Lee-Preskill algorithm. Besides the aforementioned advantages, \Han{free theory eigenstates are defined naturally under the HO basis through LSZ formalism as asymptotically far away in-states. We summarize the results in Proposition~\ref{Proposition} and explain more details in Section.\ref{Comparison} and in Appendix.}  
	
    \item Initial state preparation. In order to prepare the interaction wave packets, Jordan-Lee-Preskill algorithm uses adiabatic state preparation to turn on the coupling from free theory constructed under field basis. In the variational setup, alternative strategies could be directly used and solve the initial scattering directly. In Section \ref{initial}, we show how to prepare the state by variational algorithms. We introduce a bosonic version of the unitary coupled cluster (UCC) ansatz, which can be efficiently implemented on an NISQ device to test our simulation algorithms, and show the optimization using the imaginary time evolution~\cite{mcardle2019variational}. \Han{Our initial state preparation strategies admit a simple interpretation according to the LSZ formalism: the UCC ans\"atze act on the free far-past particle eigenstate to some wave-packet in the interaction region at $t_i$. The use of quantum imaginary time evolution further evolves this unknown wave-packet to low excited states at $t_f$. We then theoretically and numerically investigate the spectral crowding phenomena in quantum field theories in both weakly-coupled and strongly-coupled theories.}
	
	\item Real-time evolution and scattering. Besides digital quantum simulation algorithms (see Refs.~\cite{tro,low2016hamiltonian,shaw2020quantum,chakraborty2020digital,bender2018digital}), the real-time evolution algorithms could also be tracked by variational methods, see Refs.~\cite{li2017efficient,yuan2019theory,kokail2019self,Paulson_2021}. During the real time evolution, variational errors might be hard to control especially for the non-perturbative regime and violent scattering processes. Nonetheless, we can track the simulation error during the time evolution, and we can adaptively construct the quantum circuit to achieve the desired accuracy within a polynomial circuit depth \cite{zhang2020low}. A theoretical framework for the dynamics simulation and comments on the challenges of the scattering process is provided in Section~\ref{particle}. 
	
	\item  Simulation fidelity. Simulation errors in the variational setting will not only be limited to the digital simulation error (like the Trotter error) but also the variational error from the ansatz, measurement error, and noise in the devices. In this work, we observe that in the initial scattering state preparation, as long as the total particle number and type are not changed significantly, the scattering experiment could still be performed, even starting with imprecise wave packets. Thus, the task of scattering state preparation could tolerate more noise. It could be qualified by particle subspace fidelity and suitable for NISQ devices. In Section~\ref{numerics}, we show the numerical simulation for the ground state and excited states preparation using variational quantum algorithms, and compare it with adiabatic evolution by fidelities. We also provide a resource analysis of our method. Finally, in Section \ref{outlook}, we highlight a list of future directions.
\end{itemize}

We summarize error and efficiency analyses in the following  Proposition.
\sun{
\begin{proposition}\label{Proposition}
	Let us consider a discrete $\lambda \phi^4$ theory on a spatial lattice $\Omega = a \mathbb{Z}_{N}$ with total length $L$, lattice spacing $a$ and number of site $N=\frac{L}{a}$. We also define its momentum space dual lattice $\Gamma = \frac{{2\pi }}{L} \mathbb{Z}_{N}$. Let us put $n_q$ qubits at each site ($n_q = \lceil {{{\log }_2}( {1 + 2{\phi _{\max }}/{\delta _\phi }} )} \rceil$ for field basis and $n_q = \lceil \log n_{\text{cut}} \rceil$ for HO basis with more explanations on the notation in Section \ref{stateencoding}). Then the following facts hold:
	\begin{enumerate}
		\item  Let $\epsilon$ be an acceptable truncation error of truncated wavefunctions of free theory eigenstates under basis field. Then $n_q = O(\log \log \frac{1}{\epsilon})$. While eigenstates under HO basis are simulated by single computational basis elements without truncating any wavefunction. On the other hand, both simulated field operators $\phi(x), \pi(x)$ (Eq.~\eqref{FieldOperators}) and ladder operators $a_k^\dagger, a_k$  (Eq.~\eqref{LadderOperators}) do not satisfy the canonical commutation relation perfectly. Let $\epsilon'$ denote the corresponding error, then $n_q = O(\log \frac{1}{\epsilon'}))$ (see Section~\ref{Comparison} \& Appendix).
		\item The full theory Hamiltonian $H$ is sparse under HO basis with $O(N^3)$ nonzero entries for each of its row/column and can be complied by $O(N^3 n_{\text{cut}})$ Pauli operators (Section \ref{Comparison}). The UCC ansatz $\hat{T}_2$ (Eq.~\eqref{eq:UVCC_pair}) proposed in Section \ref{UCC} can be be compiled by $O(N n_{\text{cut}}^2)$ Pauli operators acting on $2n_q$ qubits. To judge our algorithm, an $n$-particle subspace fidelity can be measured in $O(N^n)$ times (see Section \ref{fidelity}).
	\end{enumerate}
\end{proposition}
}

\section{Formalism and state encoding}\label{stateencoding}

At the starting point, we show how to encode our Hamiltonian from quantum field theory to a quantum device. 
In Section.~\ref{intro}, we first give a review of the $\lambda \phi^4$ theory in 1+1 dimension, and we point out the use of harmonic oscillator basis in calculating scattering amplitudes in the scalar field theory by Lehmann-Symanzik-Zimmermann (LSZ) reduction formula. 
Then, we provide a  detailed comparison on various versions of bases, including the field basis and the harmonic oscillator basis in the coordinate space and momentum space, respectively. 
Some similar discussions can be found in the quantum chemistry context \cite{sawaya2020resource}.

\subsection{$\lambda \phi^4$ theory and the LSZ reduction formula}\label{intro}

In this theory, we have a scalar quantum field $\phi$ with the Hamiltonian 
\begin{align}
	H = \int {dx\left( {\frac{1}{2}{\pi ^2} + \frac{1}{2}{{\left( {{\partial _x}\phi } \right)}^2} + \frac{1}{2}m_0^2{\phi ^2} + \frac{{{\lambda _0}}}{{4!}}{\phi ^4}} \right)}. \nonumber
\end{align}
Moreover, we discretize it in the lattice, 
\begin{align}
	H = \sum\limits_{x \in \Omega } a \left[ {\frac{1}{2}{\pi ^2} + \frac{1}{2}{{\left( {{\nabla _a}\phi } \right)}^2} + \frac{1}{2}m_0^2{\phi ^2} + \frac{{{\lambda _0}}}{{4!}}{\phi ^4}} \right]. \nonumber
\end{align}
The theory is defined on the spatial lattice $\Omega$ with dual lattice $\Gamma$ (see Proposition \ref{Proposition}). The field momentum $\pi(x)$ is defined as the Fourier conjugate of the field $\phi({x})$ with the following commutation relation,
\begin{align}
	&\text{lattice: }[\phi (x),\pi (y)] = i{a^{ - 1}}{\delta _{x,y}},\nonumber\\
	&\text{continuum: }[\phi (x),\pi (y)] = i\delta (x - y).\nonumber
\end{align}
The discretized version of the derivative is given by
${\left( {{\nabla _a}\phi } \right)^2}(x) = {\left( \phi \left( {x + a} \right) - \phi (x) \right)^2/{a}^2}$, where $m_0$ is the (bare) mass term in the free theory, and the $\frac{{{\lambda _0}}}{{4!}}\phi {{(x)}^4}$ term represents the coupling. When $\lambda_0=0$, we call it the free theory. In the case of the free theory, we could diagonalize the Hamiltonian by the following mode decomposition in the continuum,
\begin{align} \label{mode-expansion}
	&\phi (x) = \int {\frac{{{d}k}}{{{{2\pi }}}}} \sqrt {\frac{1}{{2{\omega _k}}}} \left( {{a_k} + a_{ - k}^\dag } \right){e^{ipx}},\nonumber\\
	&\pi (x) =  - i\int {\frac{{{d}k}}{{{{2\pi }}}}} \sqrt {\frac{{{\omega _k}}}{2}} \left( {{a_k} - a_{ - k}^\dag } \right){e^{ikx}}.
\end{align}
The discrete version can be defined similarly, where the canonical algebra of $\phi$ and $\pi$ leads to the commutation relation as
$[ {{a_k},a_l^\dag } ] = L{\delta _{k,l}}$ for lattice and 
$[ {{a_k},a_l^\dag } ] = 2\pi \delta (k - l)$ for {continuum}, respectively.
The energy dispersion is given by
\begin{align}
	\omega (k) = \sqrt {m_0^2 + \frac{4}{{{a^2}}}{{\sin }^2}\left( {\frac{{ak}}{2}} \right)} \xrightarrow{a\to 0} \omega_k \equiv \sqrt {m_0^2 + {k^2}} .
\end{align}
In such a basis, the Hamiltonian is diagonalized as 
\begin{align} \label{FreeHamiltonian}
	{H_0} = \sum\limits_{k \in \Gamma } {\frac{1}{L}} \omega (k)a_k^\dag {a_k} + {E_0} \xrightarrow{a\to 0} \int {\frac{{dk}}{{2\pi }}} {\omega _k}a_k^\dag {a_k} + {E_0},
\end{align}
where
\begin{align}
	{E_{0}} = \sum\limits_{k \in \Gamma } {\frac{1}{2}} \omega (k) \xrightarrow{a\to 0}  \int {\frac{dk}{2\pi}}{\frac{1}{2}} \omega_k \times 2\pi \delta(0).
\end{align}
\Han{
	An important physical observable we could measure, is the (Wightman) two-point function as
	$
	G(x - y) = \langle \Omega |\phi (x)\phi (y)|\Omega \rangle,
	$
	where $\ket{\Omega}$ is the ground state of the theory. In the free theory case where $\lambda_0=0$, one can compute the two point function explicitly
	\begin{align}
		{G_{0}}(x - y) = \sum\limits_{k \in \Gamma } {\frac{1}{L}} \frac{1}{{2\omega (k)}}{e^{ik(x - y)}} \xrightarrow{a\to 0}   \int {\frac{{dk}}{{2\pi }}} \frac{1}{2\omega _k} e^{i k (x-y)}.
	\end{align}
The two-point function of the scalar defines the scalar mass of the theory. In the continuum limit, when we turn on the interaction $\lambda_0$, in weakly-coupled regime one could compute the correction to the mass through Feynman diagrams. The theory will experience a second-order phase transition at strong coupling, where the universal behavior belongs to the 2D Ising universality class and the use of perturbation theory is difficult. In Appendix, we address the relation between lattice models and their field theory description, emphasizing the importance of simulating quantum field theories. From a non-perturbative perspective, computing two-point functions will tell us the information about masses of particles through \emph{Källén-Lehmann spectral representation} \cite{Kallen:1952zz, Lehmann_1954}. We also review necessary backgrounds through LSZ reduction formula in Appendix that is particularly suited to the use of HO basis in this paper.}

%-------------------------------------------------------------------------------------------------------------------------------------------------------------------

\subsection{The field basis in the coordinate space}
One of the simplest considerations is the field basis. For a scalar quantum field theory discretized in a lattice $\Omega$, we could define the state decomposition
\begin{align}
|\psi \rangle  = \int_{ - \infty }^\infty  d {\phi _1} \cdots \int_{ - \infty }^\infty  d {\phi _N}\psi \left( {{\phi _1}, \ldots ,{\phi _N}} \right) \ket{{\phi _1}, \ldots ,{\phi _N}}.
\end{align}
Here, $N$ is the total number of sites. The state decomposition for an arbitrary state $\psi$ gives the above wavefunction $\psi \left( {{\phi _1}, \ldots ,{\phi _N}} \right)$, \Zimu{where we abuse the notation $\phi_i$ chosen as an arbitrary number in $\mathbb{R}$ to denote an eigenvalue of the local field operator $\phi(x_i)$. The corresponding eigenstates $\ket{\phi_i}$ form a basis of the local Hilbert space.} This definition is similar to the coordinate basis in quantum mechanics. 

Now, since we are using a quantum computer, we need to truncate the local Hilbert space. \Zimu{Moreover, we want an increment $\delta_\phi$ in discretization of the spectra of each $\phi(x_i)$ such that we do not need to choose variables in a continuous interval. The states that we are interested in is truncated and discretize as
\begin{align} \label{DiscreteFieldBasis}
	\ket{\psi_{\rm{cut}}} = \sum_{\phi_1 = -\phi_{\max}}^{\phi_{\max}} \cdots \sum_{\phi_N = -\phi_{\max}}^{\phi_{\max}} \delta_ \phi^N \psi(\phi_1,...,\phi_N) \ket{\phi_1,...,\phi_N},
\end{align}
and thus the number of qubits we need to encode at each site $x_i$ is ${n_q} = \left\lceil {{{\log }_2}\left( {1 + 2{\phi _{\max }}/{\delta _\phi }} \right)} \right\rceil$. There are bounds on $\phi_{\max}, \delta_\phi$ and $n_q$ from the scattering energy $E$ derived in Refs.~\cite{Jordan:2011ne,Jordan:2011ci} which are useful to prove the polynomial complexity of the Jordan-Lee-Preskill algorithm. However, the original bound $n_q = O(\log \frac{1}{\epsilon})$ of qubit number with respect to truncation error is not tight, we compute rigorously by several properties of Hermit-Gauss functions (eigenfuncations of quantum harmonic oscillator) and show in Appendix that this bound can be refined as $n_q = O(\log \log \frac{1}{\epsilon})$.}

\subsection{The HO basis in the coordinate space}
There is another important basis, the harmonic oscillator basis to define a digital representation of states in lattice quantum field theories. We firstly consider the following transformation,
\begin{align}
	&\phi (x) = \frac{1}{{{{\left( {2{m_x}} \right)}^{1/2}}}}\left( {{a_x} + a_x^\dag } \right),\nonumber,
\end{align}
with $\pi(x)$ being transfomred similarly. Here $m_x$ is a free parameter we could choose. Then the canonical algebra of $\phi$ and $\pi$ leads to $\left[ {{a_x},a_y^\dag } \right] = {a^{ - 1}}{\delta _{x,y}}$. Now, the creation operator $a_x^\dagger$ and its conjugate could define the number states $\ket{n_x}$ at the site $x$. \Zimu{On the lattice $\Omega$ of $N$ sites, let us say that we are mostly interested in the maximal energy level $n_\text{cut}$, so we cut the Hilbert space and define}
\begin{align}
	|{\psi _{{\rm{cut}}}}\rangle  = \sum\limits_{{n_1} = 0}^{{n_{{\rm{cut}}}}}  \ldots  \sum\limits_{{n_N} = 0}^{{n_{{\rm{cut}}}}} {\psi ({n_1}, \ldots ,{n_N})} \left| {{n_1},\ldots ,{n_N}} \right\rangle.
\end{align}

\subsection{The field basis in the momentum space}
Now we introduce the dual field basis in the momentum space. Remember that we define the dual lattice $\Gamma$ based on the spatial lattice $\Omega$. Thus, one can directly write the Hamiltonian in terms of the momentum coordinate. \Zimu{To be more specific, consider the free theory mode expansion Eq.~\eqref{mode-expansion} with Fourier transformation of $\phi(x)$
\begin{align}\label{PhiExpansion}
	{\phi_k} = \frac{1}{{\sqrt {2{\omega _k}} }}\left( {{a_k} + a_{ - k}^\dag } \right),
\end{align}
and the dual field momentum $\pi_k$ being defined similarly. Then we can discretize the interaction piece of the Hamiltonian in the momentum space by
\begin{align} \label{phi4_mom}
	H_{\text{int}} = \frac{{{\lambda _0}}}{{4!}}\frac{1}{{{L^3}}}\sum\limits_{k_1,k_2,k_3 \in \Gamma } \phi_{k_1} \phi_{k_2} \phi_{k_3} \phi_{- k_1 - k_2 - k_3}.
\end{align}
We can make a truncation on the field range in the momentum space and discretize a state like Eq.~\eqref{DiscreteFieldBasis}.}

\subsection{The HO basis in the momentum space}
Similarly, we could consider the HO basis in the momentum space. \Zimu{Under this basis, a discretized state decomposition is written as
	\begin{align}
		\ket{\psi _{\rm{cut}}} = \sum_{n_{k_i} \leq n_{\text{cut}}} \psi ({n_{k_1}}, \ldots ,{n_{k_N}}) \ket{ n_{k_1},\ldots ,n_{k_N}}.
	\end{align}
The number of qubits needed at each momentum mode $k_i$ is thus $n_q = \lceil \log n_{\text{cut}} \rceil$}. The number states $\ket{n_{k_i}}$ now are generated by the creation operator from Eq.~\eqref{mode-expansion}. The above state has a very clear physical meaning: the basis directly show the scalar particle numbers in different momenta. This also provides a good initial guess for the excited states in the interacting theory. In Section~\ref{numerics}, we discuss the particle excitations in the momentum space in more details.

\subsection{A comparison}\label{Comparison}
\Zimu{Besides a brief comparison mentioned in the Introduction, we discuss more details on discretization of scalar field under different bases. For each position site $x_i$, a truncated field operator $\phi(x_i)$ is given by Pauli Z-matrices:  
\begin{align}\label{FieldOperators}
	\phi(x_i) = \frac{\phi_{\max}}{2^{n_q}} \sum_{j = 1}^{n_q - 1} 2^j Z_j
\end{align}
with $\pi(x_i)$ being defined as the discrete Fourier transform of $\phi(x_i)$. The corresponding free theory vacuum is then the discrete Gaussian prepared by Kitaev-Webb Algorithm. It is shown in \cite{Klco:2018zqz,robert2019resource,kitaev2008wavefunction} that they can be efficiently constructed in a quantum circuit. On the other hand, the creation/annihilation operators on momentum space are defined as
\begin{align}\label{LadderOperators}
	& a_{k_i}^\dagger = \sum_{s = 0}^{n_{\text{cut}} - 1} \sqrt{s + 1} \ket{(s+1)^{k_i}} \bra{s^{k_i}}, \notag \\ 
	& a_{k_i} = \sum_{s = 1}^{n_{\text{cut}}} \sqrt{s} \ket{(s - 1)^{k_i}}\bra{s^{k_i}},
\end{align}
where $k_i \in \Gamma$ specifies a momentum mode with $\ket{s}$ being the computational basis of $n_q = \lceil \log n_{\text{cut}} \rceil$ qubits at each mode. The corresponding free theory vacuum in this case is simply $\ket{s = 0}^{k_1} \otimes \cdots \otimes \ket{s = 0}^{k_N}$ and hence can be prepared at a constant circuit depth. This is also true for other excited initial state used in our numerical simulation (e.g., Eq.~\eqref{InitialState1} \& \eqref{InitialState2}). Easily constructible initial states is the first advantage when working with HO basis. Besides, these states are taken without truncating any wavefunction and hence there is no need to consider the truncation error. Even though, we should mention that the aforementioned discrete commutation relations false for both truncated energy level of HO basis and truncated field strength of field basis. The corresponding error only decays exponentially with $n_q$. More details can be seen in Appendix. }

\Zimu{Furthermore, as the computational basis encodes particle numbers of momentum modes, the free theory Hamiltonian Eq.~\eqref{FreeHamiltonian} is automatically diagonalized. To check the implementation efficiency of interaction Hamiltonian Eq.~\eqref{phi4_mom}, we first examine the sparsity. By Eq.~\eqref{PhiExpansion}, each row/column of the matrix representation of $\phi_{k_i}$ contains at most $2$ nonzero entries. Even $H_{\text{int}}$ is non-local, any of its terms $\phi_{k_1} \phi_{k_2} \phi_{k_3} \phi_{- k_1 - k_2 - k_3}$ is a four-fold tensor product and hence contains at most $2^4$ nonzero entries in each row/column. Comparing with the $2^{n_q N}$-dimensional total Hilbert space, each term is sparse. Because $H_{\text{int}}$ has $N^3$ terms when summing over the dual lattice $\Gamma$ with momentum conservation, nonzero matrix elements scales cubically with the number $N$ of momentum mode. This makes $H$ applicable under most existing quantum algorithm, especially the imaginary time evolution employed in this work.~\footnote{The Hamiltonian with truncated energy levels is represented in a low-energy subspace.} To count the number of Pauli operators needed to compile this Harmonization, we first consider how to expand creation/annihilation operators by the Pauli basis $\{I,\sigma_x, \sigma_y, \sigma_z \}^{\otimes n_q}$ at each mode. To calculate the needed number, we transform the Pauli basis into matrix unit basis $\{E_{ij}\}$: consider the $n_q = 1$ (one-qubit) case:
\begin{align*}
	& 2E_{11} = I + \sigma_z, \quad 2E_{12} = \sigma_x + i\sigma_y, \\
	& 2E_{22} = I - \sigma_z, \quad 2E_{21} = \sigma_x - i\sigma_y.
\end{align*} 
Each matrix unit $E_{ij}$ can be written as a linear combination of $2$ Pauli operators. Hence the corresponding transformation matrix $M$ has $2$ nonzero entries in each of its columns. Since the creation operator $a^\dagger$ on one-qubit space is colinear with $E_{12}$, it decomposes into $2$ Pauli operators. For two-qubit space, the transformation matrix simply equals $M^{\otimes 2}$ with $2^2$ nonzero terms. As $a^\dagger$ is now a linear combination of $E_{12}, E_{23}, E_{34}$ and one can check that $E_{12}, E_{34}$ are transformed from the same sub-collection of Pauli operators, $a^\dagger$ decomposes into $2 \cdot 2^2$ pieces in total. By induction, the decomposition of $a^\dagger$ on $n_q$-qubit space has $n_q \cdot 2^{n_q}$ pieces. With respect to the energy cut-off, we need $n_{\text{cut}} + \log n_{\text{cut}}$ Pauli operators which scales linearly with $n_{\text{cut}}$. By the same method, the Hermitian operator $a + a^\dagger$ can be built by $\frac{1}{2}(n_{\text{cut}} + \log n_{\text{cut}}$ Pauli operators with the factor $\frac{1}{2}$ coming from cancellation of anti-Hermitian terms when we sum $a$ and $a^\dagger$ together. As a simple example to verify this point, let $n_q = 1$, then $a + a^\dagger$ is colinear with a single Pauli operator $\sigma_x$. On the other hand, by Eq.~\eqref{phi4_mom} \& \eqref{PhiExpansion}, $H_{\text{int}}$ is built by at most $8 N^3 (n_{\text{cut}} + \log n_{\text{cut}})$ Pauli operators. We will apply this method to verify gates efficiency of the UCC variational ansatz in Section \ref{initial} and Section \ref{numerics}.}
	
\Han{One can check that the above analysis automatically holds for general $d+1$ dimensional theory where $N$ stands for the number of momentum nodes in $d$ dimension. Except the efficiency of preparing initial states and implementing the Hamiltonian, the HO basis is also useful to keep track of the simulation results in real-time, since one could quickly identify the basis overlap and find the particle number and their momenta. Indeed, the harmonic basis specifies the momentum sectors of the asymptotically far past in-states, where the fields satisfy the on-shell condition. For the interacting theory, when the interaction is turned on, one could specify the momentum sectors again by the adiabatic state preparation and we could use this method to \emph{define} the wave packets in the given momentum sectors from adiabetically preparing $a_F^{\dagger}(t)$, as shown in Appendix. Thus, in this paper, we will mainly work on the HO basis in the variational setup. }

\section{Variational quantum algorithms}\label{initial}
\subsection{The variational ansatz}
Variational quantum simulation is a useful technique especially for the near-term quantum computer. The variational algorithm starts by preparing the quantum state by a quantum circuit as
\begin{align}
\left| {\psi ( \theta )} \right\rangle  = \left( {\prod\nolimits_{\ell  = 1}^L {{U_\ell }\left( {{\theta _\ell }} \right)} } \right)\left| {\psi_{0}} \right\rangle.
\end{align}
Here $U_{\ell}$s are some unitary operators that could be realized in the quantum device, for instance,
$
{U_\ell }\left( {{\theta _\ell }} \right) = {e^{-i{\theta _\ell }{X_\ell }}},
$
with the variational parameters  $\theta = (\theta_\ell)$. $X_\ell$s are some Hermitian operators, for instance, elements in the Pauli group, and $\ket{\psi_0}$ could be some simple initial states that could be easily prepared. The target state will, in principle, be approximated by some optimal choices of ${\theta}$, say ${\theta}^*$, which could be found using the variational principles. For example, a typical problem in quantum simulation is to find the ground state, then we could minimize the energy with respect to the variational parameters $\langle H \rangle_{\theta} \equiv   \left\langle \psi ( \theta )\right|H\left| {\psi ( \theta )} \right\rangle$.

\Han{
The general strategy for the ground state searching is by updating the parameters as
\begin{align}
{\theta _\mu }(t + 1) = {\theta _\mu }(t) - \sum_{\nu}{\eta _\mu }(t) A^{-1}_{\mu \nu}(\theta(t)) \frac{\partial }{{\partial {\theta _\nu }}}{\langle H \rangle_{\theta(t)}},
\label{eq:update}
\end{align}
where ${\theta _\mu }(t)$ represents the optimization dynamics with step $t$, and the learning rate is given by $\eta_\mu(t)$. Here, we use $A(\theta(t))$ to represent the metric matrix at the parameter $\theta(t)$. }The metric matrix in the gradient descent algorithm is simply the identity matrix. In the following section, we will show its explicit form during the optimization. \Han{One can ask if there exists a regime where there is a convergence guarantee and, if so, the rate of convergence for these variational paramterization. One can study this question from over-paramterization using quantum neural tangent kernel (QNTK) \cite{liu2022analytic}. Further taking $\langle H \rangle_{\theta(t)} \equiv z(\theta(t))$, Eq.\eqref{eq:update} implies:
\begin{align}
    \begin{aligned}
    &z(\theta(t + 1)) - z(\theta(t)) \equiv \delta z = \sum \frac{\partial z}{\partial \theta_{\mu}} \delta \theta \\
    &= - \sum_{\nu} A^{-1}_{\mu \nu}(\theta(t)) \eta_{\nu}(t)\sum_{\mu} \frac{\partial z(\theta(t))}{\partial \theta_{\mu} } \frac{\partial z(\theta(t))}{\partial \theta_{\nu}} \\
    \end{aligned}
\end{align}
Assuming $A$ is identity matrix and $\eta$ to be paramterization-independent, the resultant is precisely the QNTK defined in \cite{liu2021representation}. Note that we can interpret $A^{-1}_{\mu \nu}$ as the \emph{learning rate tensor} as part of definition of NTK in classical neural networks \cite{robert2019resource}.  In particular, for the circuits that form at least approximate 2-design that satisfies certain concentration conditions (See in \cite{liu2022analytic}), the average convergence is of the form
\begin{align}
    \epsilon(\theta(t)) \approx e^{- \gamma t}\epsilon(\theta(0)),
\end{align}
where $\epsilon \equiv z(\theta) - E_0$,  $E_0$ is the ground state energy and
\begin{align}
    \gamma \approx \frac{ \eta L \operatorname{tr}(H^2)}{\operatorname{dim}(\mathcal{H})^2},
\end{align}
with $L$ being the total count of variational parameters and $H$ be full Hamiltonian. The dimension of the Hilbert space in our case is $n_{\operatorname{cut}}^N$. The exponential convergence rate is guarantee on average in the \emph{over-parametrization regime} where $L \approx \operatorname{dim}(\mathcal{H})^2 /\operatorname{tr}(H^2) $. 
When $A$ fails to be an identity matrix such as in the case of quantum imaginary time evolution used in the following, no precise analytical convergence guarantee is known. However, it seems to be reasonable to extrapolate the hypothesis that such methods, due to its more physical/geometric nature, would have convergence rates lowered-bounded by the naive gradient descent methods. The dependence of the square of size of Hilbert space would imply the above analysis only is suitable to small size qubit system (See in Appendix for more details). 
}

The next question is how to choose the initial state $\ket{\psi_{0}}$ and $U_{\ell}$s? The precise strategy of choosing $\ket{\psi_{0}}$, $U_{\ell}$ and the optimization scheme will specify the variational quantum algorithm we use. There are many variational algorithms (see Ref.~\cite{sugurureview21,cerezo_variational_2021} for a recent review). In this work we will discuss the following bosonic unitary coupled cluster (UCC) ansatz and imaginary time evolution where we practically find the best in our physical system. Different from the quantum computational chemistry literature, where the UCC ansatz consists of the fermionic excitations in the active space, our algorithm expresses the UCC ansatz directly with the bosonic mode. 

%-------------------------------------------------------------------------------------------------------------------------------------------------------------------
\subsection{Bosonic UCC ansatz}\label{UCC}
	
\Zimu{As is mentioned before, the variational algorithm may not be very sensitive to the locality of the Hamiltonian. With known implementation efficiency, we will focus on the HO basis in the momentum space.} Prior work has extensively investigated the coupled cluster methods to solve the electronic energy spectra and vibrational structure in the chemistry and materials science, and the quantum version, unitary coupled cluster ansatz, has been suggested and further experimentally demonstrated to solve the chemistry problems on a quantum computer~\cite{O_Malley_2016,Shen_2017}. Other prior works~\cite{C9SC01313J,ollitrault2020hardware} discussed the usage of bosonic UCC in studying vibronic properties of molecules.

The general form of unitary coupled cluster is given by 
\begin{equation}
	\ket{\psi(\theta)} = \exp(i\hat T) \ket{\psi_0},
\end{equation}
where $\hat {T}$ is the sum of symmetry preserved excitation Hermitian operators truncated at finite excitations as $\hat T = \hat {T}_1 + \hat {T}_2+\cdots$. The key ingredient of UCC ansatz is to search for the true ground state of the interacting fermionic theory by considering the particle-conserving excitations above a reference state.

In our quantum field theory setup, a bosonic version of the UCC ansatz \cite{C9SC01313J,ollitrault2020hardware} could be natural to capture types and particle numbers for scalar particles. In the momentum space, the effective action preserves the momentum reflection symmetry ($k \to -k = \tilde k$). \Zimu{Therefore, we may express the $l$th excitation Hermitian operators of the bosonic UCC ansatz in the momentum space as
\begin{align} \label{eq:UVCC}
	\hat{T}_\ell = \sum_{k_1,...,k_\ell} & \sum_{|s_i - t_i| \leq 4} \theta_{s_1^{(k_1)}, t_1^{(k_1)},...,s_\ell^{(k_\ell)}, t_\ell^{(k_\ell)} } \times \notag \\
	& \Big(\ket{ s_1^{(k_1)},...,s_\ell^{(k_\ell)}} \bra{ t_1^{(k_1)},...,t_\ell^{(k_\ell)}} \notag \\
	& + \ket{ s_1^{(\tilde{k}_1)},...,s_\ell^{(\tilde{k}_\ell)}} \bra{ t_1^{(\tilde{k}_1)},...,t_\ell^{(\tilde{k}_\ell)}} \Big).
\end{align}	
Here, $k_1,...,k_\ell$ are $\ell$ distinct momentum modes taken from the lattice $\Gamma$. Since the $\lambda \phi^4$ field could lift the excitation up to 4 level, we impose the energy constraint $|s_i - t_i| \leq 4$ for $\hat{T}_\ell$ at each momentum mode which makes each term of $\hat{T}_\ell$ spares like the $\lambda \phi^4$ Hamiltonian. To count Pauli operators, we first note that the concerned local Hilbert space is defined by $\ell n_q$ qubits (recall that $n_q = \lceil \log n_{\text{cut}} \rceil$) and then apply the same from Section \ref{Comparison} to $\hat{T}_\ell$. With energy constraint, it can be built by $O(n_{\text{cut}}^\ell)$ Pauli operators and each of which acts nontrivially on at most $\ell n_q$ qubits. The total ansatz is thus compiled by $O(N^\ell n_{\text{cut}}^\ell)$ Pauli operators with the same order of number of parameters. When $\ell = 4$, expanding $H_{\text{int}}$ by Eq.~ \eqref{phi4_mom}, \eqref{PhiExpansion} \& \eqref{LadderOperators}, we can set parameters of $\hat{T}_4$ being the expansion coefficients. We can even vary these parameters as $\theta(s), s \in [0,1]$ such that $H_{\text{int}} = \hat{T}_4 ( \theta(1))$ and hence
\begin{align}\label{eq:turn_on}
	H(s) = H_0 + s\hat{T}_4 ( \theta(t) )
\end{align}
is tantamount to adiabatic turn-on of the interaction. To simulate the adiabatic evolution, we have to divide the time interval $[0,1]$ into $M$ pieces with $M$ large enough (depending on the energy gap). We then apply Trotter formula to approximate the time evolution using $O(N^4 n_{\text{cut}}^4 )$ Pauli operators for each product term.}

\Zimu{In NISQ devices however, we wish to further reduce the computational cost. Thus we focus on variational ans\"atze and restrict to use single excitation operator $\hat{T}_1$ which can be constructed by at most $4N \frac{1}{2}(n_{\text{cut}} + \log n_{\text{cut}})$ Pauli operators. The second term is obtained like expanding $a + a^\dagger$ in Section \ref{Comparison}. We also employ a modified double excitation operators $\hat{T}_2$ as
\begin{align}
	\hat{T}_2 = & \sum_{k} \sum_{|s_i - t_i| \leq 4} \theta_{s^{(k)}_{1},  t^{(k)}_{1},s^{(\tilde k)}_{2}, t^{(\tilde k)}_{ 2}}  \times\nonumber\\
	&\left(\ket{s^{(k)}_{1}  s^{(\tilde k)}_{2}}\bra{t^{(k)}_{1} t^{(\tilde k)}_{ 2}} +\ket{s^{(\tilde k)}_{1}  s^{(k)}_{2}}\bra{t^{(\tilde k)}_{1} t^{k)}_{ 2}} \right),
	\label{eq:UVCC_pair}
\end{align}
which considers the pairing correlations of the momentum $k$ and $\tilde k$. It makes $k$ the only momentum variable when taking summation. With the requirement to be Hermitian and the energy constraint, this ansatz can be built by at most $32N \frac{1}{2}(n_{\text{cut}} + \log n_{\text{cut}}$ Pauli operators such that each of which acts nontrivially on at most $2n_q$ qubits and hence reduces a large number of parameters comparing Eq.~\eqref{eq:UVCC}. We may even discard the second term in Eq.~\eqref{eq:UVCC_pair} to further reduce the gate count in the variational quantum circuits. For small $n_{\text{cut}}$, the number of used Pauli operators would be even fewer (see Fig.~\ref{fig:pauli_term} in Section \ref{numerics}). As an inevitable consequence, these variational ans\"atze cannot replace adiabatic evolution in searching ground state. We will remedy this problem by employing the quantum imaginary time evolution in the next section.}

%-------------------------------------------------------------------------------------------------------------------------------------------------------------------
\subsection{Variational state preparation}
We now discuss how to use variational quantum algorithms for finding the ground state and the low-lying excited states.
We first briefly review the variational quantum simulation algorithm of imaginary time evolution~\cite{mcardle2019variational}. The normalized imaginary time evolution at imaginary time $\tau$ is given by 
$
\left| {\psi (\tau )} \right\rangle  = \frac{{{e^{ - H\tau }}\left| {{\psi _0}} \right\rangle }}{{\sqrt {\left\langle {{\psi _0}} \right|{e^{ - 2H\tau }}\left| {{\psi _0}} \right\rangle } }}.
$
The population of the energy eigenstate $\ket{e_j}$ will decay exponentially with the energy $E_j$, and the ground state can be obtained in the long time limit 
$
\ket{\psi^{(0)}}=\lim_{\tau \rightarrow \infty} \ket{\psi(\tau)}.
$
While the nonunitary imaginary time evolution cannot be directly implemented on a quantum computer, one could still simulate imaginary time evolution on a quantum computer by using the hybrid quantum-classical algorithm. Instead of simulating the imaginary time evolution directly, we assume that the time-evolved state can be approximated by a parametrized trial state 
$
\ket{\psi( \theta (\tau))},
$
with variational parameters 
$
\theta (\tau)=(\theta_\mu (\tau)).
$
As mentioned in \cite{mcardle2019variational}, by minimizing the distance between the ideal evolution and the evolution of the parametrized trial state, the evolution of the target state $\ket{\psi(\tau)}$ under the Schr\"{o}dinger equation can be mapped to the trial state manifold as the evolution of parameters $\theta$. 

Using McLachlan's variational principle, we have \begin{align}
	\delta \left\| {\left( {{{d_{\tau }}} + H - {E_\tau }} \right)\left| {\psi (\theta (\tau ))} \right\rangle } \right\| = 0,
	\label{eq:ITE_principle}
\end{align}
and the evolution of the parameters under the imaginary time evolution could be determined by 
\begin{align}
	\sum_{j} A_{i, j} \dot{\theta}_{j}=-C_{i},
\end{align}
with the matrix elements of $A$ and $C$ given by
\begin{align}
	A_{i, j} &=\text{Re} \left( {\partial_{i}}\langle\psi(\theta(\tau))| {\partial_{j}}|\psi(\theta(\tau))\rangle \right),\nonumber\\
	C_{i} &=\text{Re}  \left( {\partial_{i}}\langle\psi(\theta(\tau))| H|\psi(\theta(\tau))\rangle \right).
	\label{eq:AandC}
\end{align}
Here, $\||\psi\rangle \|=\sqrt{\langle\psi | \psi\rangle}$ is the norm of the quantum state, we denote ${\partial_{i}} \equiv \partial/{\partial{\theta_i}}$, and we assume the parameters are real.
By tracking the evolution of the variational parameters, we can effectively simulate imaginary time evolution.
This actually serves as an optimizer to update the parameters in Eq.~\eqref{eq:update}.
It is worth mentioning for the pure state imaginary time evolution, this approach is equivalently to the quantum natural gradient descent method, and the matrix $A$ is indeed the Fisher matrix \cite{stokes2019quantum}. 

The quantum imaginary time evolution minimizes the energy loss function:
\begin{align}
	\mathcal{L}(\vec{\theta}(\tau)) =  \frac{1}{2} \left( \bra{\psi(\vec{\theta}(\tau))} H \ket{\psi(\vec{\theta}(\tau))} - E_0 \right)^2 \equiv \frac{1}{2}\varepsilon^2,
\end{align}
where its total variation \cite{mcardle2019variational}: 
\begin{align}
	\begin{aligned}
		\frac{d\mathcal{L}(\vec{\theta}(\tau)) }{d \tau} &= \varepsilon \operatorname{Re}(\bra{\psi(\theta(\tau))}H \frac{\ket{\theta(\tau)}}{d\tau}) \\
		&= - \varepsilon \sum_{ij}C_i A^{-1}_{ij} C_j \\
		& \leq 0, 
	\end{aligned}
\end{align}
where the fact that the $\sum_{ij}C_i A^{-1}_{ij} C_j$ is nonnegative follows that the fact that $A$ is non-negative definite. The nonnegativity of $\varepsilon$ follows from the variational theorem where $E_0$ is the smallest eigenvalue. 

Moreover, having found the ground state $\ket{{\psi^{(0)}}}$, we can construct a modified Hamiltonian 
$
H^{(1)}=H+\alpha \ket{\psi^{(0)}}\bra{\psi^{(0)}},
$
where $\alpha$ is the regularization term that lifts the ground state energy, and is sufficiently large comparing to the energy scale of the system. 

The ground state of the modified Hamiltonian $H^{(1)}$ becomes $\ket{\psi^{(1)}}$, the first excited state $\ket{\psi^{(1)}}$ of the original Hamiltonian $H$. As $\ket{\psi^{(0)}}$ is  an excited state of the modified Hamiltonian, we can evolve the system under $H^{(1)}$ in the imaginary time to suppress the spectral weight of $\ket{\psi^{(0)}}$ and  obtain the first excited state $\ket{\psi^{(1)}}$. This process can be repeated to obtain the higher-order  excited states. More specifically, the ($n+1$)th excited state is the ground state of effective Hamiltonian 
\begin{align}\label{eq:hamiltonian-adpative}
	H^{(n+1)}=H+ \alpha \sum_{j=0}^n |{\psi^{(j)}}\rangle \langle{\psi^{(j)}}|.
\end{align}
Instead of preparing the Hamiltonian directly, we can simulate the imaginary time evolution under $H^{(n+1)}$ by tracking the evolution of the parameters, which are now modified as 
\begin{align}
	C_{i} =&\text{Re} ( {\partial_{i}}\langle\psi(\theta(\tau))| H|\psi(\theta(\tau))\rangle + \nonumber\\
	&\alpha \sum_{j=0}^n {\partial_{i}}\langle\psi(\theta(\tau)) |{\psi^{(j)}}\rangle \langle{\psi^{(j)}}|\psi(\theta(\tau))\rangle ),
\end{align}
while the matrix $A$ keeps the same as in Equation (\ref{eq:AandC}).
These addition terms in $C_i$ can be evaluated using the low-depth swap test circuit.  Other variational excited state preparation techniques can be found in a recent review paper~\cite{sugurureview21}.

We wish to remark that the circuit ansatz for the imaginary time evolution does not have to be fixed. Instead, the circuit ansatz could be adaptively determined by tracking the distance of the ideal evolved state and the variational state. In the extreme case, we could construct the circuit by approximating the normalized state at every single time step, which reduces to the quantum imaginary time evolution, firstly proposed in Ref.~\cite{motta2020determining}. Suppose the Hamiltonian has the decomposition  
	$
	H = \sum_{l=1}^L \hat{h}_l ,
	$
	where the Hamiltonian contains $L$ local terms and each $\hat h_l$ acts on at most $k$ neighboring qubits.
	Using the first-order Trotterization, the evolved state after  applying nonunitary operator $e^{-\delta \tau \hat h_l }$ within imaginary time $\delta \tau$ by  
	\begin{equation}
		\left|{\Psi (\tau + \delta \tau )}\right\rangle = c^{-1 / 2} e^{-\delta \tau \hat h_l }|\Psi(\tau)\rangle \approx e^{-i \delta \tau \hat{A}}|\Psi(\tau)\rangle,
		\label{eq:QITE}
	\end{equation}
	where $c$ is the normalization factor and $\hat{A}$ is a Hermitian operator that acts on a domain of $D$ qubits around the support of $\hat h_l$.
	The unitary operator $e^{-i \delta \tau \hat{A}}$ can be determined by minimizing the approximation error in Eq.~\eqref{eq:QITE}, which is similar to the derivation in Eq.~\eqref{eq:ITE_principle}. For a nearest-neighbor local Hamiltonian on a $d$-dimensional cubic
	lattice, the domain size $D$ is bounded by $\mathcal{O} (C^d)$, where $C$ is the correlation length. More details about the algorithm complexity can be found in Ref.~\cite{motta2020determining}.
	
	This circuit construction strategy can be regarded as a special case in the variational imaginary time evolution given by Eq.~\eqref{eq:ITE_principle} and Eq.~\eqref{eq:AandC}.
	If we fix the old circuit ansatz $\theta(\tau)$ constructed before imaginary time $\tau$, and determine the new added unitary operator $\theta(\delta \tau)$ that approximates the effect of $e^{-iH \delta \tau}$, this is exactly the same as  Eq.~\eqref{eq:QITE}.
	However, to further reduce the circuit depth, we can jointly optimize the parameters 
	$
	\theta(\tau) \oplus \theta(\delta \tau),
	$
	making it more compatible for the near-term quantum devices.

%-----------------------------------------------------------------------------------------------------------------------------------------------------
\subsection{Spectral crowding}
Before we start to apply variational algorithms, we will make a short investigation on the spectrum of the $\lambda \phi^4$ quantum field theory. In the momentum space, HO basis, one might have a large number of degeneracies in the energy eigenstates (similar problems appear in other bases as well), bringing potential problems for quantum simulation. We will borrow the terminology ``spectral crowding" that has been used in the ion trap systems \cite{landsman2019two} referring to this situation. 

\begin{figure}[tb]
 \centering
 \includegraphics[width=0.82\linewidth]{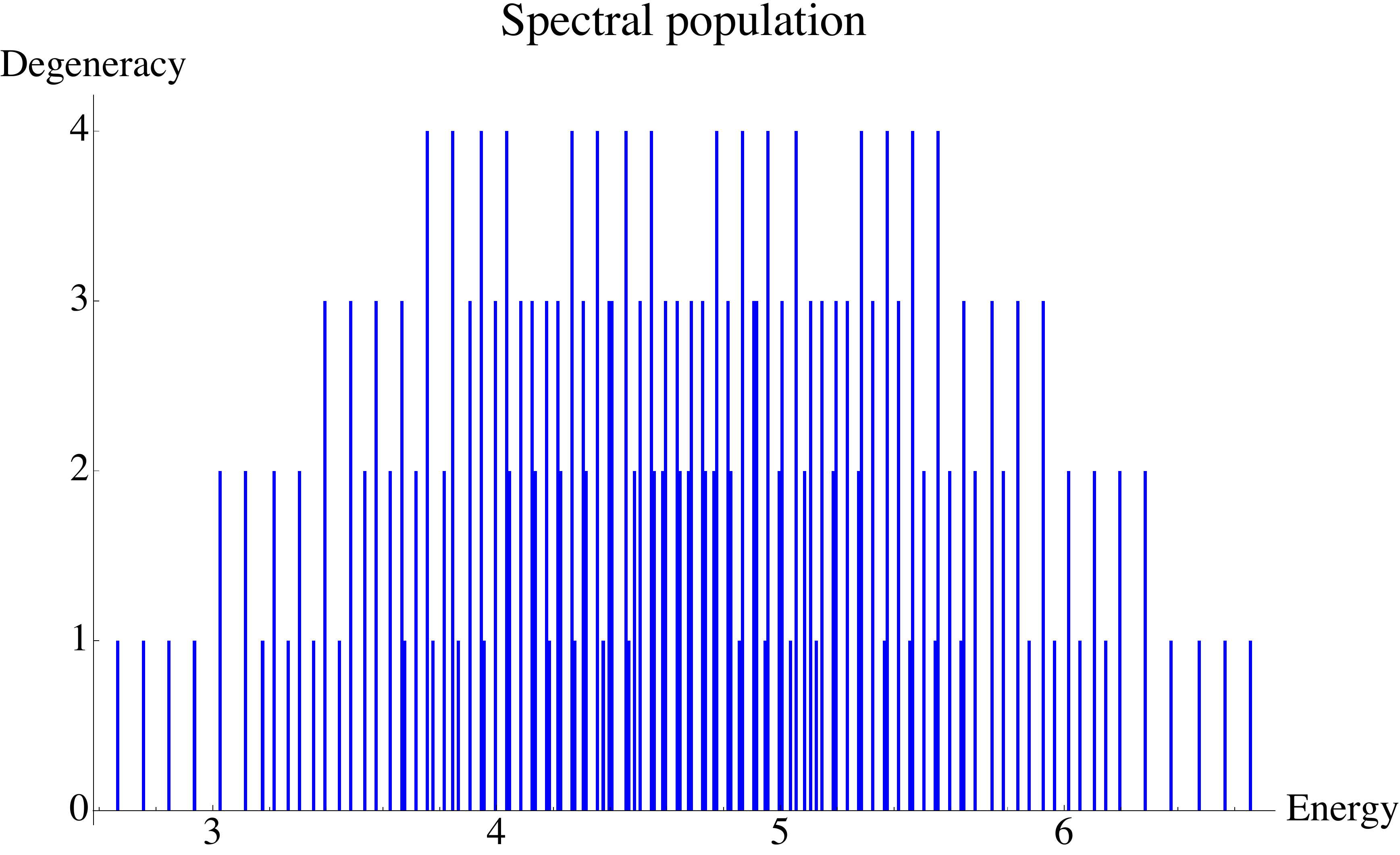}
 \includegraphics[width=0.82\linewidth]{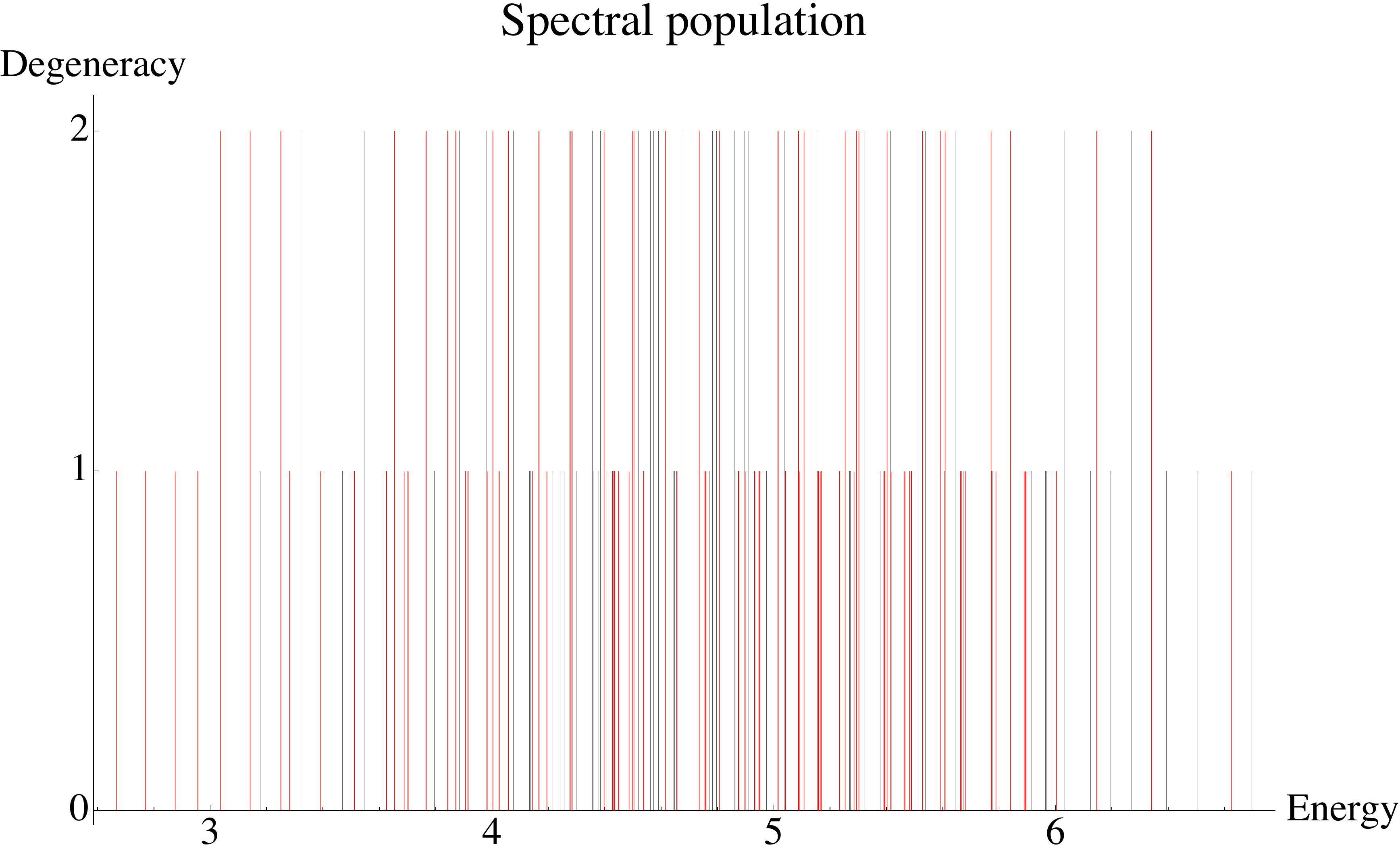}
 \caption{Spectral crowding for $\lambda \phi^4$ theory: Blue/left: free theory; Red/right: interacting theory. We use $m_0=0.369$ and $\lambda_0=0$ (free) or $\lambda_0=0.481$ (interacting), with maximally three excitations $n_{\text{cut}}=3$, system size $N=4$, and the lattice spacing $a=1$, for the HO basis in the momentum space.
 }
 \label{fig:spec}
 \end{figure}

For excited states, degeneracy might happen even in the free theory in our construction. For instance, say that in the free theory, it might be the case where
\begin{align}
\sum\limits_i {{n_i}\omega ({p_i})}  = \sum\limits_j {{{\bar n}_j}\omega ({{\bar p}_j})}.
\end{align}
Here, we have states represented in the HO basis in the momentum space, with particle numbers and momenta $n_i, p_i$, or $\bar{n}_j, \bar{p}_j$, and their energies are precisely identical. A typical example is that considering the continuum limit, we might have
\begin{align}
n\left| {{m_0}} \right| = \sqrt {m_0^2 + {p^2}},
\label{eq:identical}
\end{align} 
where $n\in \mathbb{Z}_{> 0}$. In those cases, their states are degenerate. Another typical case the role of parity which anti-commutes with the momentum:
\begin{align}
\omega (p) = \omega ( - p),
\end{align}
since we are not able to distinguish the left-moving and right-moving states only by their energies. Figure \ref{fig:spec} provides an example for spectral crowding, where we fix $m_0=0.369$ and $\lambda_0=0$ (free) or $\lambda_0=0.481$ (interacting), with maximally three excitations $n_{\text{cut}}=3$, system size $N=4$, and the lattice spacing $a=1$. The choice of parameters is aiming on avoiding the situation in the Eq.~\eqref{eq:identical}.

Spectral crowding might bring us difficulties on identifying states in the output, and defining different directions of momenta for particles, especially when states are excited. Instead of looking at the general structure of density of states, we start with the maximally one-particle states in this simple system. In the free theory, we have the ground state with the energy 2.662. Moreover, the single-particle states have the energies:
\begin{align}
p=\frac{2\pi}{L}(0,1,2,3): E=2.754, 3.027, 3.170, 3.027.
\end{align}
We know that this degeneracy is made by the boundary condition of the momentum $p \sim 2\pi/L - p$, which is the parity $\mathbb{Z}_2$~\footnote{However, for this set of parameter choices, the energies of two-particle and three-particle zero-momentum states are lower than the single-particle excited states}. 
In general, for an $n$-particle state, since we could freely choose the direction of momentum, the spectral crowding will be enhanced at least $O(2^n)$.

\begin{figure}[tb]
 \centering
 \includegraphics[width=0.8\linewidth]{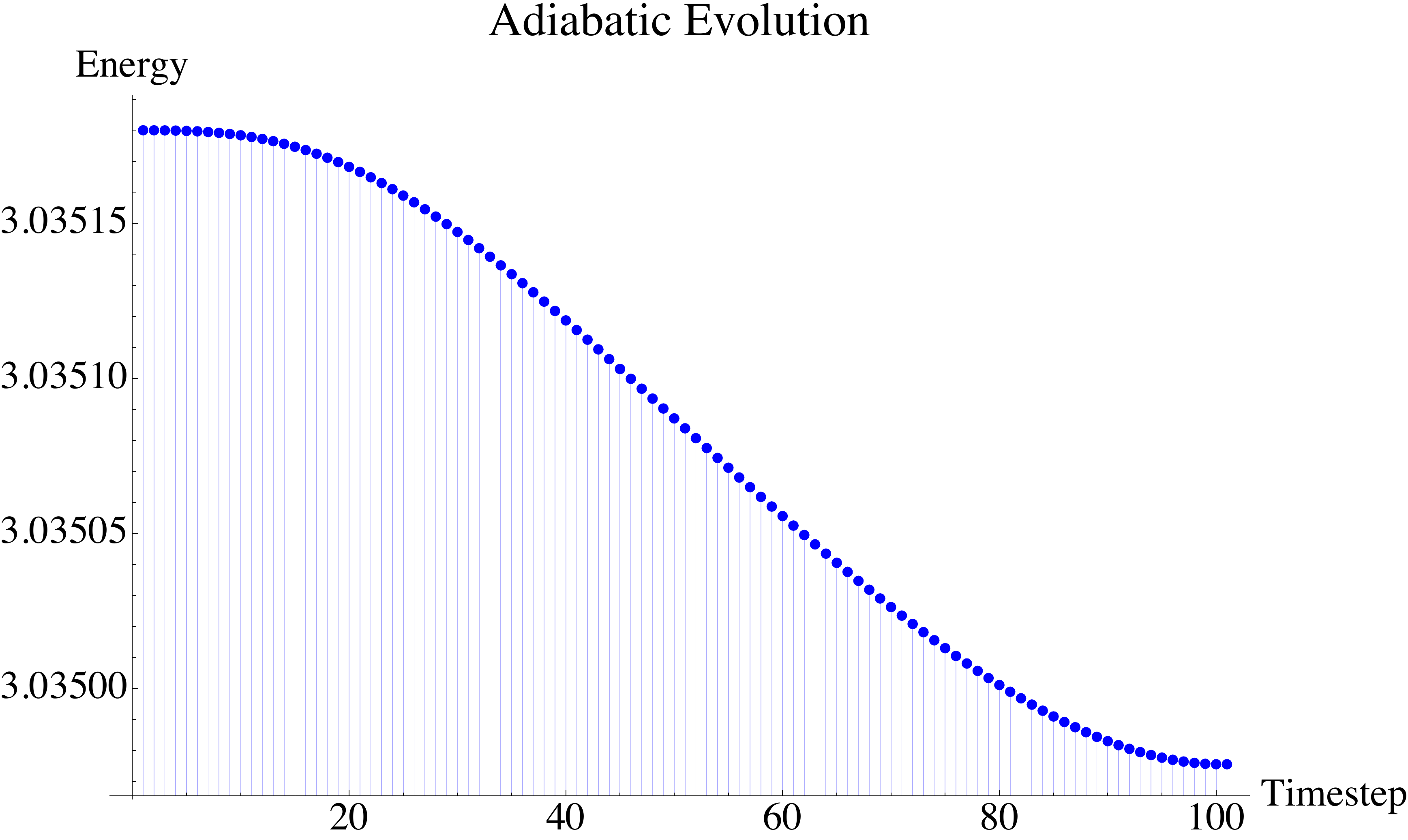}
 \includegraphics[width=0.8\linewidth]{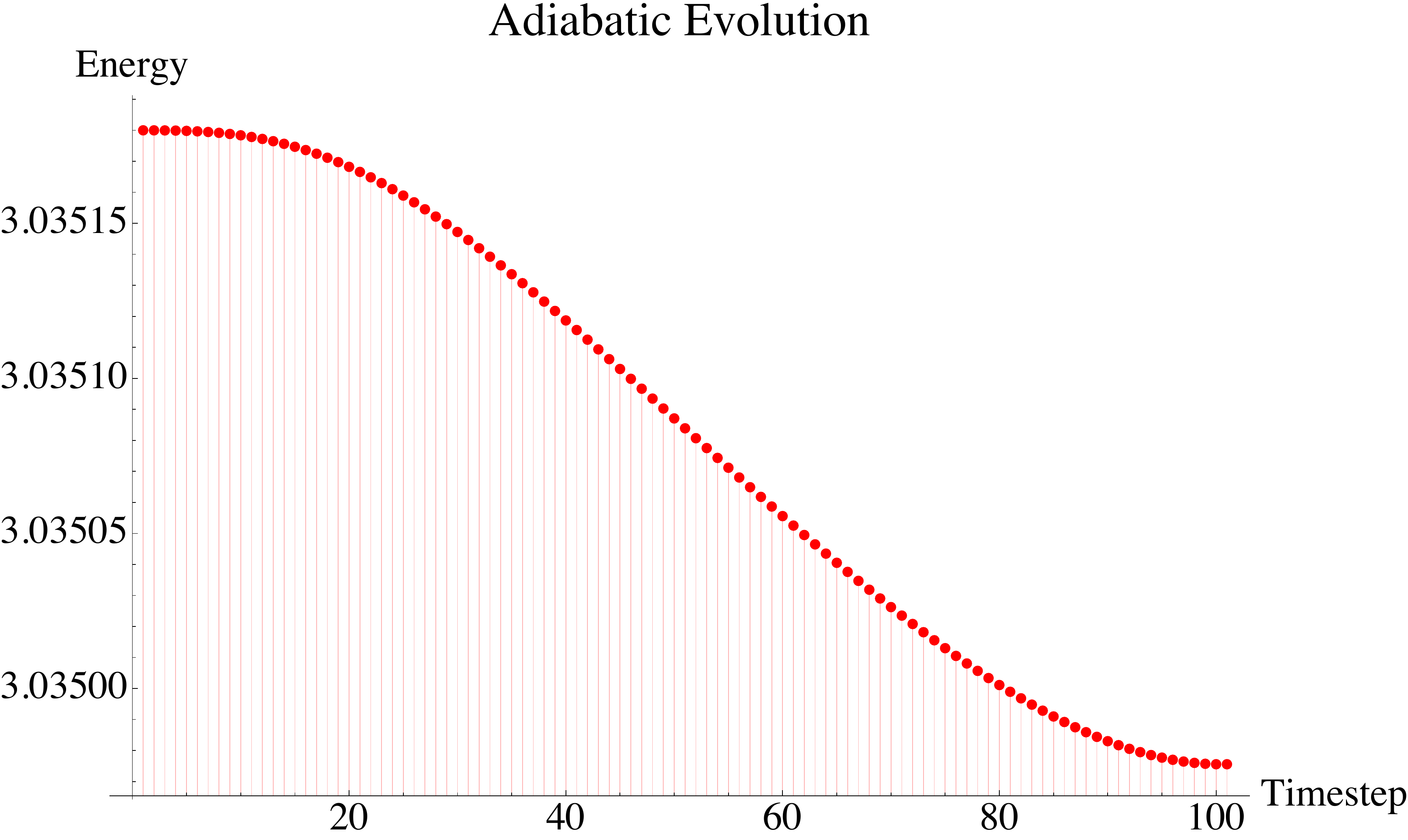}
 \caption{Adiabatic evolution starting from the free particle momenta $p=\frac{2\pi}{L}(1,3)$ (left/blue,right/red). Those two examples have the adiabatic errors both around $0.07\%$.
 }
 \label{fig:adiapro}
 \end{figure}

Now, we consider to turn on the interaction. In the adiabatic process where we slowly turn on the $\phi^4$ Hamiltonian as Eq.~\eqref{eq:turn_on}. Since the interacting Hamiltonian is invariant under the parity transformation, we could use the adiabatic process to \emph{define} the direction of the momentum. In Figure \ref{fig:adiapro}, we show an example of the adiabatic evolution numerically, with the number of adiabatic steps $T=100$ (which means that we are dividing the interval $s\in [0,1]$ to 101 steps). We find all single-particle eigenstates could agree with the corresponding energy eigenstates with high fidelities (we only show $p=\frac{2\pi}{L}(1,3)$ example in the plot, but all four adiabatic state preparations are also verified). Note that this operation specifies the direction of momenta in the interacting theory. This is an advantage of our basis, where we could specify the direction of momenta in this way. 

The above algorithm could also be made variationally. Recall that in the variational process, we are starting from a wave packet state $\ket{\psi}_\text{initial}$, and we slowly turn on the interaction $\lambda$ from the free theory $\lambda=0$. Thus, during this process, the Hamiltonian is time-dependent. Instead of considering Lie-Trotter-Suzuki decomposition in a digital quantum computer, one could perform the above calculation in a quantum computer with a variational form. We will use the variational approach of time evolution introduced in Refs.~\cite{yuan2019theory,mcardle2019variational}. Similar to the imaginary time proposal, we will use the McLachlan's variational principle and take care of the time-dependent global phase. 

Now consider the situation where we adiabatically turn on the coupling of the Hamiltonian. We restrict our solution inside the variational form similar from before,
\begin{align}
\left| {\psi (\theta )} \right\rangle  = \left( {\prod\nolimits_{\ell  = 1}^L {{U_\ell }\left( {{\theta _\ell }} \right)} } \right)\left| {{\psi _{{\text{free theory states}}}}} \right\rangle.
\end{align}
Note that we are starting from the corresponding momentum eigenstates of the free particle. The differential equation of $\theta$ based on the McLachlan's variational principle is given by
\begin{align}
\sum\limits_j {{M_{i,j}}} \frac{{d{\theta _j}}}{{ds}} = {V_i},
\label{eq:theta_dynamics}
\end{align}
where
\begin{align}
&{M_{i,j}} = {\mathop{\rm Re}\nolimits} {A_{i,j}} + {\partial_{i}}\langle\psi(\theta(s))|\psi (\theta (s))\rangle {\partial_{j}}\langle\psi(\theta(s))|\psi (\theta (s))\rangle, ~\nonumber\\
&{V_i} = {\mathop{\rm Im}\nolimits} {C_{i}} + i{\partial_{i}}\langle\psi(\theta(s))|\psi (\theta (s))\rangle \langle \psi (\theta (s))|H(s)|\psi (\theta (s))\rangle,~\nonumber
% &{A_{i,j}} = {\partial_{i}}\langle\psi(\theta(s))|{\partial_j |\psi (\theta (s))\rangle },\nonumber\\
% &{C_i} = {\partial_{i}}\langle\psi(\theta(s))|H(s)|\psi (\theta (s))\rangle.
\end{align}
and $A$ and $C$ are similarly defined in Eq.~\eqref{eq:AandC}.
One could show that the solution of $\theta$s are always real in our variational form, and the variational answer is consistent with the actual answer up to a time-dependent global phase. More detailed discussions on the simulation error during the dynamics can be found in Section~\ref{particle}.

Similar to the above variational adiabatic state preparation algorithm, the imaginary time evolution could also start from the corresponding free theory states. Practically, we find in our example, the imaginary time evolution algorithm performs better (this is intuitively because we are looking for low-lying states with low energies). 

Moreover, these methods can be integrated together. We can turn on the interaction, similarly as in Eq.~\eqref{eq:turn_on}, but with much fewer steps. We could then use the variational algorithms to find the ground state of the intermediate Hamiltonian $H(s)$, using which as the initial state for the next time step until $s$ reaches $1$. Compared to finding the ground state of $H_I$, this method may avoid the local minimum.

Finally, we comment on other methods to snake around the spectral crowding problem. A useful trick to find the state with both fixed momentum and energy is through measurement in quantum devices. We could consider measuring the momentum operator
\begin{align}
P = a\sum\limits_{x \in \Omega } {\pi {\nabla _a}\phi },
\end{align}
during the variational process, making sure that it keeps the sign when the interaction is turning on. However, the momentum operator only has its meaning in the free theory, so we only expect the above algorithm to be useful in the sense of weakly-coupled theory. 

Another useful trick for keeping the momentum is similar to the idea of the tangent space method in the language of matrix product state (see a review \cite{vanderstraeten2019tangent}). Usually, we expect that our momentum$-p$ eigenstate could have the following form
\begin{align}
\left| {{\Phi _p}} \right\rangle  = \sum\limits_{x \in \Omega } {{e^{ipx}}{T_x}\left| \Phi  \right\rangle }.
\end{align}
Here $T_x$ is the translation operator with the vector $x$. If the state $\ket{\Phi}$ is already a momentum-$p$ eigenstate, 
\begin{align}
\left| \Phi  \right\rangle  = \sum\limits_{y\in \Omega} {{e^{ipy}}{T_y}\left| \Psi  \right\rangle },
\end{align}
we have
\begin{align}
&\sum\limits_{x \in \Omega } {{e^{ipx}}{T_x}\sum\limits_{y \in \Omega } {{e^{ipy}}{T_y}\left| \Psi  \right\rangle } }  = \sum\limits_{x,y \in \Omega } {{e^{ip(x + y)}}{T_{x + y}}\left| \Psi  \right\rangle }  \nonumber\\
&= \sum\limits_{x,z \in \Omega } {{e^{ipz}}{T_z}\left| \Psi  \right\rangle }  \propto \sum\limits_{z \in \Omega } {{e^{ipz}}{T_z}\left| \Psi  \right\rangle }.
\end{align}
Moreover, if the state $\ket{\Phi}$ is a linear superposition of the momentum-$p$ state and the momentum-$(-p)$ state where $p\ne 0$
\begin{align}
&\sum\limits_{x \in \Omega } {{e^{ipx}}{T_x}\left( {{c_1}\sum\limits_{y \in \Omega } {{e^{ipy}}{T_y}\left| \Psi  \right\rangle }  + {c_2}\sum\limits_{y \in \Omega } {{e^{ - ipy}}{T_y}\left| \Psi  \right\rangle } } \right)} \nonumber\\
&= \#  \times {c_1}\sum\limits_{z \in \Omega } {{e^{ipz}}{T_z}\left| \Psi  \right\rangle }  + {c_2}\sum\limits_{x,y \in \Omega } {{e^{ip(x - y)}}{T_{x + y}}\left| \Psi  \right\rangle }  \nonumber\\
&= \#  \times {c_1}\sum\limits_{z\in \Omega} {{e^{ipz}}{T_z}\left| \Psi  \right\rangle }  + {c_2}\sum\limits_u {{e^{ipu}}} \sum\limits_v {{T_v}\left| \Psi  \right\rangle } \nonumber\\
&\sim \#  \times {c_1}\sum\limits_{z \in \Omega} {{e^{ipz}}{T_z}\left| \Psi  \right\rangle }.
\end{align} 
Note that the $c_2$ term is suppressed because it sums over a pure numerical phase. Thus, for the state we obtained from the variational quantum simulation, we could make a linear superposition weighted by $e^{i p x}$ to obtain a momentum eigenstate with a fixed momentum direction, at least in the case of the single-particle scattering experiment. However, the above method seems to be mostly useful when we know how to construct the translation operator. It is manifest in the coordinate space, but not easy in the momentum space.

\subsection{State fidelity, one-particle subspace fidelity and generalizations}\label{fidelity}
Here we discuss some concepts about fidelities that are useful for the variational, scattering-state preparation setting. Say that we originally have a wave packet centered around a given momentum, and it is a one-particle state in the free theory. Now we could turn on the interaction slowly. Ideally, as we discussed before, a one-particle state will still remain a one-particle state in the interacting theory. In fact, if we consider momentum eigenstates of a single particle, $\ket{p}$, we could define the one-particle subspace by
\begin{align}
{V_{\text{one-particle,free}}} = {\text{span}_{{p}}}(\ket{p}).
\end{align}
Now, if we are adiabatically turning on each state $\ket{p}$ towards the coupling $\lambda_0$, the space will become
\begin{align}
{V_{\text{one-particle,}\lambda_0}} = {\text{span}_{{p}}}(\text{adiabatic evolution}_{\lambda_0} \circ \ket{p}).
\end{align}
In fact, if the adiabatic evolution is slow enough, the above expression will define the one-particle space in the interacting theory. \Zimu{This makes our number eigenstate definition more precise. Counting the one-particle eigenstates in free theory on the lattice $\Omega$, they span an $N$-dimensional subspace. The dimension of a $n$-particle space is equal to the number of \emph{compositions} $(n_1,...,n_N)$ of $n$. It is mathematically analogous to calculate the dimension of $n$th symmetry tensor power of $\mathbb{R}^N$ and the answer is $\binom{n+N-1}{n}$. It should be noted that since we also truncate particle numbers by $n_{\text{cut}}$ at each momentum mode, an n-particle space with $n > n_{\text{cut}}$ cannot be fully realized and thus has a lower dimension. Even though, low energy subspaces can be fully defined with dimension being bounded polynomially in $N$.}

\begin{figure*}[tbp]
\centering
\includegraphics[width=0.88\linewidth]{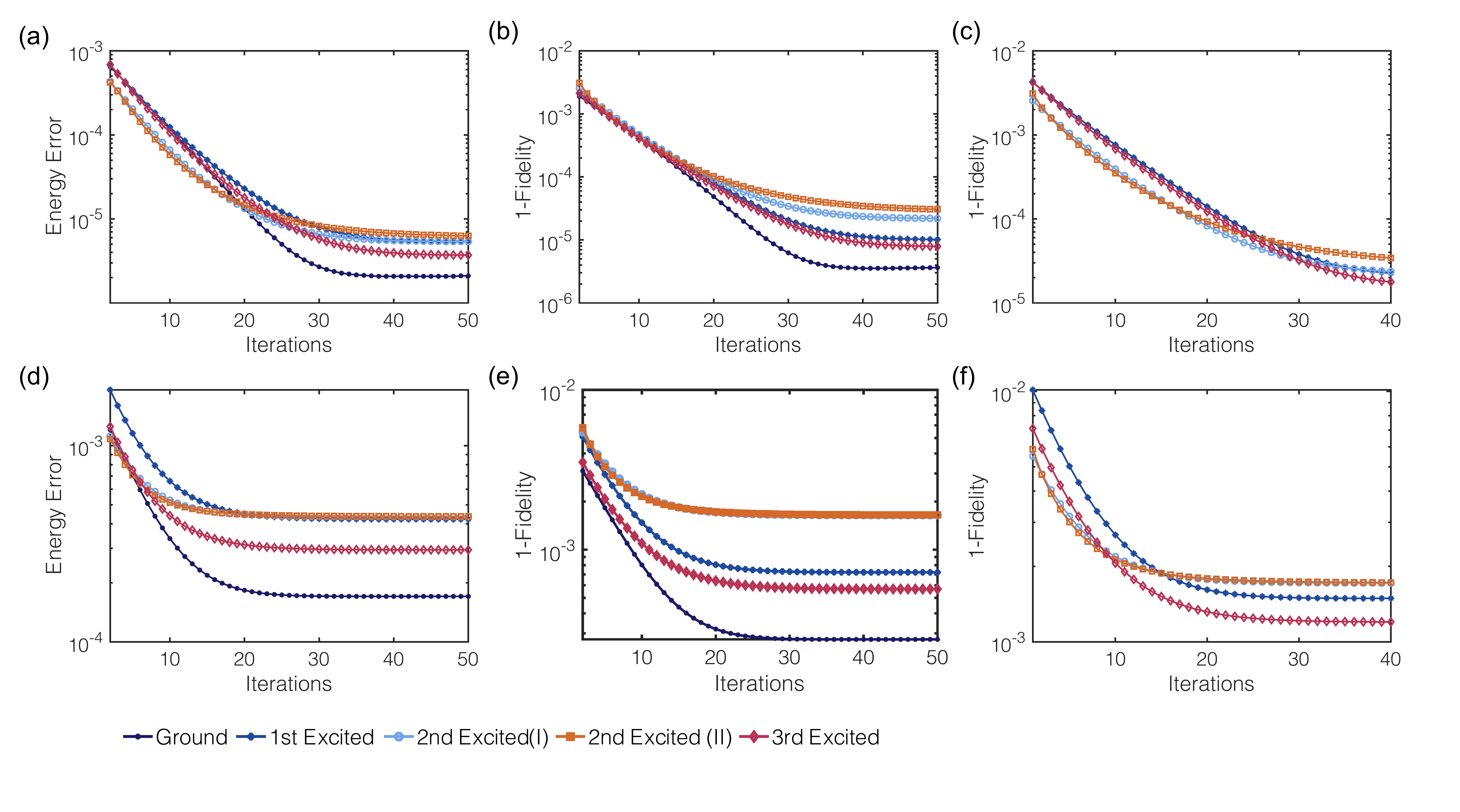}
\caption{ 
Convergence towards the ground state and excited states of $\lambda \phi^4$ theory in the momentum space using variational algorithms. 
The static mass is set to be $m_0=1.27$. Figure (a-c) corresponds to $\lambda_0 = 1$  and Figure (d-f) corresponds to $\lambda_0 = 10$, respectively.
The initial state for the excited states searching is the corresponding single-particle excited states of the free Hamiltonian $\lambda_0 = 0$.
The error of the results with exact diagonalization.
(a,d) We show the relative energy error of the ground state and the low-lying excited states. 
(b,e) We show the fidelity error of the ideal eigenstates and the simulated variational states.
(c,f) The fidelity error of the first four excited states in a one-particle subspace.
}
\label{fig:mom_m1p3}
\end{figure*}

Now, say that we are doing the state preparation using the variational algorithm (which is not the ideal adiabatic process). Due to the limitation induced by the variational ansatz, we will have some systematic errors (or other errors). However, the resulting state, although suffering from the noise, might still have a large overlap with the one-particle subspace ${V_{\text{one-particle,}\lambda_0}}$. In fact, we could define the state fidelity
\begin{align}
{F_{{\rm{state,adiabatic}}}} = \left| {\left\langle {{\psi _{{\rm{ideal}}}}|{\psi _{{\rm{variational}}}}} \right\rangle } \right|,
\end{align}
which is an overlap between the accurate state from an ideal adiabatic simulation without any error, and the state obtained from the variational algorithm. We could also define the one-particle subspace fidelity
\begin{align}
{F_{{\text{one-particle,adiabatic}}}} = \left| {\left\langle {{\psi _{{\rm{variational}}}}|\Lambda |{\psi _{{\rm{variational}}}}} \right\rangle } \right|.
\end{align}
Here, $\Lambda$ is the projector of the space ${V_{\text{one-particle,}\lambda_0}} $. \Zimu{By definition, ${F_{{\text{one-particle,adiabatic}}}}$ should be no less than ${F_{{\rm{state,adiabatic}}}}$ (e.g., Fig.~\ref{fig:mom_m1p3} $(b),(c)$ or $(e),(f)$). In principle, we wish our fidelities to be always high enough, which means that we are performing high-quality state preparations. Ideally, we wish the state fidelity to be high. But in practice, when we do not really care about the form of the wave packet, and we only care about if the state is still approximately a one-particle state, we could only use the one-particle subspace fidelity. As a summary, the level of requirements we want about fidelities is closely related to the actual physical motivation we have in the simulation experiment.}

Let us end this subsection by making a final comment on the fidelities. The definition of fidelities is indeed related to the physical task we want when doing the experiment. The definition of one-particle subspace fidelity corresponds to the choice when we wish to maintain the one-particle subspace during the initial state preparation. When we have other requirements, we could demand other versions of fidelities be high. For instance, we could define the momentum fidelity by measuring the momentum center of the wave packet. We could also define the wave packet profile fidelity by measuring the wave packet form. Stronger definitions on fidelities would require higher quality when we are doing the variational state preparation.

In the most general setting, we could define the projector to the subspace $\Lambda$ as 
\begin{align}\label{projector}
\Lambda  = \sum\limits_{i = 1}^{{D_\Lambda }} {\left| {{q_i}} \right\rangle \left\langle {{q_i}} \right|}, 
\end{align}
where $\ket{q_i}$ is the basis and $D_{\Lambda}$ is the dimension of the subspace. Thus, for a variational state $\ket{\psi}$ we have
\begin{align}
&\left| \psi  \right\rangle  = \sum\limits_{i = 1}^{{D_\Lambda }} {{c_i}\left| {{q_i}} \right\rangle }  + {c_e}\left| e \right\rangle  = \Lambda \left| \psi  \right\rangle  + {c_e}\left| e \right\rangle  \nonumber\\
&= \Lambda \left| \psi  \right\rangle  + (1 - \Lambda )\left| \psi  \right\rangle.
\end{align}
Here $c_i$ and $c_e$ are the expansion coefficients, and $\ket{e}$ is the perpendicular component of the subspace projector $\Lambda$. Say that the state is normalized, we have
\begin{align}
\left( {\sum\limits_{i = 1}^{{D_\Lambda }} {c_i^*{c_i}} } \right) + {\left| {{c_e}} \right|^2} = 1.
\end{align}
\begin{align}
{F_\Lambda } = \left| {\left\langle \psi  \right|\Lambda \left| \psi  \right\rangle } \right| = \sum\limits_{i = 1}^{{D_\Lambda }} {c_i^*{c_i}}  = 1 - {\left| {{c_e}} \right|^2}.
\end{align}
At the same time, the ideal state is given by the expansion coefficients $d_i =c_i+\varepsilon_i$, 
$
\left| {{\psi _{{\rm{ideal}}}}} \right\rangle  = \sum\limits_{i = 1}^{{D_\Lambda }} {{d_i}\left| {{q_i}} \right\rangle }
$, and thus the state fidelity is given by
\begin{align}
&{F_{{\rm{state}}}} = \left| {\left\langle {{\psi _{{\rm{ideal}}}}|\psi } \right\rangle } \right| = \left| {\sum\limits_{i = 1}^{{D_\Lambda }} {d_i^*{c_i}} } \right|\nonumber\\
&= \left| {\sum\limits_{i = 1}^{{D_\Lambda }} {c_i^*{c_i}}  + \sum\limits_{i = 1}^{{D_\Lambda }} {\varepsilon _i^*{c_i}} } \right| = \left| {1 - {{\left| {{c_e}} \right|}^2} + \sum\limits_{i = 1}^{{D_\Lambda }} {\varepsilon _i^*{c_i}} } \right|\nonumber\\
&= \left| {{F_\Lambda } + \sum\limits_{i = 1}^{{D_\Lambda }} {\varepsilon _i^*{c_i}} } \right|.
\end{align}
This equation illustrates the relation between the state fidelity and the subspace fidelity. In the limit where $\varepsilon$s are small, those two fidelities are almost equal~\footnote{There are some recent research about quantum simulation in the low energy subspace, see Ref.~\cite{csahinouglu2021hamiltonian}. }. \Zimu{We verify numerically in Section \ref{numerics} a high one-particle subspace fidelity (Fig.~\ref{fig:mom_m1p3}) as well as state fidelity (Fig.~\ref{fig:mom_m0p5_1}, \ref{fig:mom_m0p5_10} \ref{fig:mom_m0p37_1} \& \ref{fig:mom_m0p37_10}) for a lower number of clean qubits as a successful benchmark for our variational algorithm about the adiabatic state preparation before particle scattering. State fidelities can be easily measured by Hadamard test with the help of adiabatic quantum computing. For $n$-particle subspace fidelities, as we analyzed above, the projector $\Lambda$ defined in Eq.~\eqref{projector} can be built by at most $\binom{n+N-1}{n}$ eigenstates evolving adiabatically from the free theory. Thus measuring $O(N^n)$ times like the case of state fidelity, we can compute the subspace fidelity. Especially for $n = 1$, one particle subspace fidelity can be obtained by $N$-time measurements.}

\subsection{Particle scattering}\label{particle}
Similar to the variational state preparation, we could also make the variational version of particle scattering. Now, when we are considering the variational time evolution, the only difference comparing to the adiabatic state preparation, is that now the Hamiltonian is static, and the bare coupling $\lambda_0$ is fixed. The evolution of the variational parameter is given by
\begin{align}
&\sum\limits_j {{M_{i,j}}} \frac{{d{\theta _j}}}{{dt}} = {V_i}
\label{eq:theta_dynamics2}
\end{align}
where $M$ and $V$ can be similarly expressed as
\begin{align}
&{M_{i,j}} = {\mathop{\rm Re}\nolimits} {A_{i,j}} + {{\partial_i \langle \psi (\theta (t))}}|\psi (\theta (t))\rangle {{\partial_j \langle \psi (\theta (t))|}}|\psi (\theta (t))\rangle,\nonumber\\
&{V_i} = {\mathop{\rm Im}\nolimits} {C_{i}} + i{{\partial_i \langle \psi (\theta (t))}}|\psi (\theta (t))\rangle \langle \psi (\theta (t))|H|\psi (\theta (t))\rangle.\nonumber
% \label{eq:theta_dynamics2}
\end{align}
Here the Hamiltonian $H$ does not depend on the time $t$.

Now we show how to approximate the ideal evolved state during the dynamics up to given error $\varepsilon$.
Suppose at time $t$ the ideal state $\Phi(t)$ is approximated by 
$\Phi(t) \approx \ket{\Psi(\theta(t))}$.
Within time step $\delta t$, we evolved the state $\ket{\Psi(\theta(t))}$ by updating the parameters $\theta(t)$ to $\theta(t+\delta t)$, which introduces an approximation error at $t+\delta t$ as
\begin{equation}
   \delta \varepsilon = \| \ket{\Psi(\theta(t+\delta t))} - e^{-iH \delta t} \ket{\Psi(\theta(t)}\|.
\end{equation}
Minimizing the error will give the similar results as that from the McLachlan's variational principle in Eq.~\eqref{eq:theta_dynamics2}.
In the extreme case, if we choose the unitary operators in the ansatz $U_{\ell }$ from all the Hamiltonian terms $(\hat{h}_l)$, for instance the Trotterization,  this error could be reduced to zero.
This indicates that for single step, we can guarantee the simulation error  at each time $t$ up to certain threshold.

We can also track the accumulated error during the whole scattering process. Starting from an initial state $\ket{\Psi_0}$,  the accumulated error until time $t+\delta t$ can be bounded  by
\begin{equation}
\begin{aligned}
\varepsilon &=  \| \ket{\Psi(\theta(t+\delta t))} - e^{-iH \delta t} \ket{\Phi(t)}\|\\
& \leq \sum_{\delta t} \|  \ket{\Psi(\theta(t+\delta t))} - e^{-iH\delta t} \ket{\Psi(\theta(t)))}  \| =\sum_{\delta t} \delta \varepsilon,
\end{aligned}
\end{equation}
where we have used triangle inequality and the distance invariance under the unitary transformation in the second line.
The single step error can be bounded by
\begin{equation}
\delta \varepsilon  =\sqrt{\Delta^2 \delta t^{2}+O\left(\delta t^{3}\right)},
\end{equation}
where the first-order order error is
\begin{equation}
\Delta^2 =\left\langle H^{2}\right\rangle+\sum_{j j^{\prime}} A_{j j^{\prime}} \dot{\theta}_{j} \dot{\theta}_{j^{\prime}}-2 \sum_{j} C_{j} \dot{\theta}_{j},
\end{equation}
with the matrix $A$ and $C$ defined in Eq.~\eqref{eq:AandC}.
The total error during the simulation can be bounded by
\begin{equation}
  \varepsilon \leq  T\max \Delta,
\end{equation}
where $\max \Delta$ is the  maximum error during the evolution.
In practice, we could add the operators from the Hamiltonian term $(\hat h_l)$ into the circuits to decrease the error to a certain threshold $\varepsilon_0$ by setting $\Delta_{\textrm{cut}} = \varepsilon_0/T$. Therefore, by tracking the simulation error at each step, we can ensure the simulation accuracy.

If at time $t$, the error $\Delta(t)$ is measured to be above the threshold, i.e., $\Delta(t) > \Delta_{\textrm{cut}}$, we repeat to add new operators from  the Hamiltonian term $(\hat h_l)$ until $\Delta \leq \Delta_{\textrm{cut}}$
The efficiency of the adaptive strategy is guaranteed by the following theorem.
\begin{theorem}
[Theorem 1 in Ref.~\cite{zhang2020low}].
\begin{enumerate}
\item  The first-order error $\Delta$ strictly decreases at each iteration until $0$;
\item  In each circuit construction process (if $\Delta(t) > \Delta_{\textrm{cut}}$), each Pauli term, $\hat{h}_l$, in the Hamiltonian is only needed to appear once;
\item We can achieve an error $\Delta\leq\Delta_{\operatorname{cut}}$ in at most $L$ iteration for any $\Delta_{\operatorname{cut}}\geq 0$. Here, $L$ is the number of terms in the Hamiltonian
\end{enumerate}
\label{thm:n_iter_bound}
\end{theorem}
The key idea of the proof is that in the  circuit construction subroutine, there always exists an operation $\hat{h}_k \in (\hat{h}_l)$, by appending which to the
old circuit, the distance strictly decreases if $\Delta \neq 0$. Theorem~\ref{thm:n_iter_bound} indicates that circuit construction process will terminate in a finite
number of steps during the total time evolution. In an extreme case, we can  optimize the parameters directly to make $\Delta \leq\Delta_{\operatorname{cut}}$, such that no additional gates are required to be added.  This reduces to the conventional variational algorithms in Eq.~\eqref{eq:theta_dynamics2}.

We also remark that $\Delta$ is a measurable quantity. The additional measurement cost for the adaptive circuit construction comes from $\langle{H^2}\rangle$, which could be measured efficiently by using the compatibility of the Pauli operators in the Hamiltonian. For instance, if $\hat h_l$ and $\hat h_k$ qubit-wise commute with each other, we can simultaneously measure them within one Pauli basis, which can significantly reduce the measurement cost.

We conclude this section by making the following comments about the variational realization of the particle scattering algorithm.  
\begin{itemize}
\item The bosonic ansatz Eq.~\eqref{eq:UVCC} in the momentum space allows the creation of new particles during the scattering process, and they could capture the particle excitations along time evolution.
\item Since we are considering the scattering process of the wave packet states, some challenges might appear because of the limitation of the variational ansatz: we cannot cover the full space during the variational simulation. Furthermore, besides the error appearing in the near-term quantum devices, we might have some other errors in the variational process due to the level crossing phenomena among different excited states. For a given theory, lots of tests need to be done to obtain some numerically satisfying results, and we leave those opportunities to future research. 
\item During the scattering process, we might wish to read off some explicit results for the S matrix elements. Thus, the result should be sensitive to the error, from the adiabatic state preparation to the scattering process. In this situation, we don't want uncontrolled errors from the quantum noise or some systematic errors from the assumption of the variational ansatz. However, if we only want some collective, statistical properties of the output states, for instance, some macroscopic quantities or random averages that could contain some intrinsic noises (for instance, the jets), we might have fewer constraints on the fidelity of the variational algorithm. 
\item Other hybrid-classical quantum simulation methods, such as hybrid tensor networks~\cite{yuan2020quantum}, could be leveraged to simulate this scattering process with fewer quantum resources. Moreover, perturbative quantum simulation methods that do not rely on the circuit ansatz could be applied to this task~\cite{sun2021perturbative}.
\end{itemize}

\begin{figure}[tbp]
\centering
\includegraphics[width=1.05\linewidth]{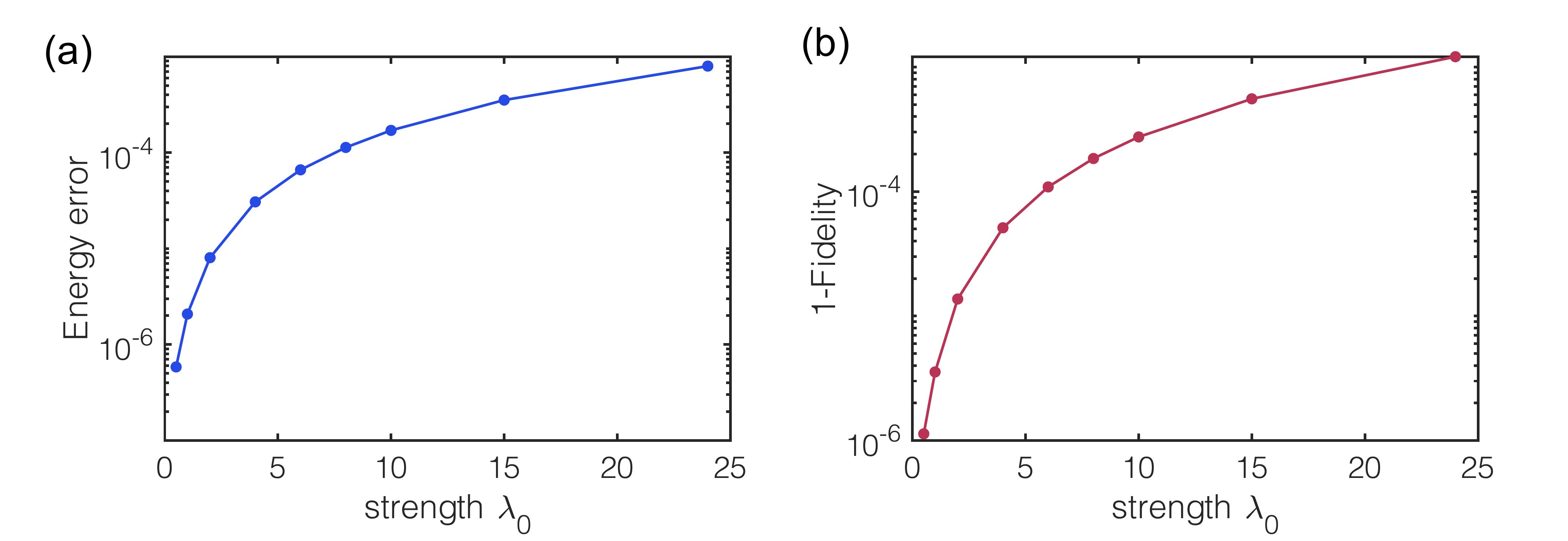}
\caption{ 
Variational ground state preparation in the momentum space. The initial state is prepared as $\ket{0}^{\otimes 8}$ in the computational basis, i.e., the vacuum state of the free Hamiltonian.
(a) The error of the ground state energy with an increasing strength of the $\phi^4$ field $\lambda_0$.  (b) The fidelity error of the ideal ground state and the variational state with an increasing strength of the $\phi^4$ field  $\lambda_0$.
}
\label{fig:lam0}
\end{figure}

\section{Numerical results}
\label{numerics}
In this section, we demonstrate how the techniques can be used to find the ground state and excited states of the interacting lattice field. We also discuss the spectrum of the lattice field with different bare mass and coupling strength. 

Similar to the analysis before, we consider a lattice $\Omega$ with total length $L=4$ and lattice spacing $a=1$, and its dual lattice $\Gamma$ has $4$ sites in the momentum space. We use the HO basis in the momentum space with $n_{\text{cut}}=4$ energy levels. To benchmark the performance of the variational algorithms, we consider finding the ground state using the bosonic UCC ansatz with an increasing coupling of the interacting field.  We prepare the initial state as the ground state, $\ket{0}^{\otimes 8}$, in the free theory. Here, we use the compact mapping for the creation and annihilation operators in Eq.~\eqref{LadderOperators}. We truncate the highest energy level to be $3$ in the double excitation operator in Eq.~\eqref{eq:UVCC_pair} to reduce the number of parameters and  the quantum circuit depth.
In order to find the variational state, we use the imaginary time evolution to evolve and identify the low-lying spectra in the interacting theory. The regularization term in the excited state search is fixed as $\alpha = 8$. In the numerical simulation, the error of the results is compared with exact diagonalization.

The relative error of the ground state energy and the state associated with different coupling strengths of the interacting field, $\lambda_0$,  is shown in Figure~\ref{fig:lam0}. 
We characterize the relative energy error by
$(E_{\textrm{variational}} - E_{\textrm{ideal}})/ E_{\textrm{ideal}}$, where $E_{\textrm{ideal}}$ is the correpsonding eigenenergies calculated by the exact diagonalization.
The state fidelity is defined by the overlap between the variational state and the ideal excited state $\left\langle {{\psi _{{\rm{ideal}}}}|{\psi _{{\rm{variational}}}}} \right\rangle $.
We can see that the simulation error increases when the interaction strength $\lambda_0$ increases, ranging from $\lambda_0 = 0.5 \sim 4!$, but even for a large interaction strength $\lambda_0 = 4!$, we can achieve a high simulation accuracy below $10^{-3}$ both for the energy error and the state fidelity, which indicates a strong representation capability of the quantum circuit ansatz.  In the following, we will choose two coupling strength $\lambda_0 = 1$ and $\lambda_0 = 10$ to test the performance of the variational algorithms in several regimes~\footnote{For sufficiently small $\lambda_0$, the eigenstates in the free theory is close to that in the interacting theory.}.

Moreover, we extend the discussions to the excited state preparation, which could be more complicated due to the spectral crowding and degeneracy of the lattice field.
We first consider the static mass $m_0=1.27$, such that the energy of single-particle excitation is lower than that of multi-particle excitations in both the free field and interacting field.
In this case, we prepare the initial state for the excited states, searching in the corresponding single-particle excited-state space of the free  Hamiltonian $\lambda_0=0$.

We show the relative error of the energy and the fidelity of the ground state and low-lying excited states towards the iteration, see Figure \ref{fig:mom_m1p3}, (a-c) for $\lambda_0=1$, and (d-f) for $\lambda_0=10$, respectively. 
As is noticed before, the single-particle excitation has a two-fold degeneracy for excitation at momentum $p=2\pi/L (1,3)$ due to the boundary condition of the momentum. For the degenerate states, we compare the state fidelity in the subspace of the degenerate states. From the simulation results, we can find that the eigenstates obtained from the variational algorithms can be found with a high state fidelity, verifying the effectiveness of the variational algorithms.
 
Figure \ref{fig:mom_m1p3}, (c,f) shows
the state fidelity in the one-particle subspace.  Here, the one-particle subspace is obtained by adiabatically evolving the one-particle state in the free theory.  In the adiabatic evolution, we set the time step $dt = 0.01$ and total time $T=50$ to ensure the state fidelity error below $10^{-4}$. 
The results indicate a high state overlap in this one-particle space in the presence of interaction $\lambda_0 = 1$ and $\lambda_0 = 10$, consistent with our analysis.

We then discuss the simulation in the spectral crowding regime with a relatively small static mass $m_0$, where the many-particle state occupied at zero momentum $p=0$ will have lower energies compared to the single-particle state. In this regime, 
we should note that the excitations in the interacting field will not be local and may not have a well-defined particle number as that in the free theory.
Therefore, searching for the excited state could be difficult in general, even when we could have access to adiabatic evolution. 
In what follows, we will discuss the low-lying excited states for two static mass $m_0 = 0.5$ and $m_0 = 0.37$, and show the search for eigenstates using the variational algorithms. 

Here, we represent the state as 
\begin{equation}
    \ket{\psi_0} = \ket{n_0, n_1, n_2, n_3},
\end{equation}
where $n_j$ denotes the  occupation number at the momentum $p = \frac{2\pi j}{L}$.

\begin{figure}[htbp]
\centering
\includegraphics[width=1.04\linewidth]{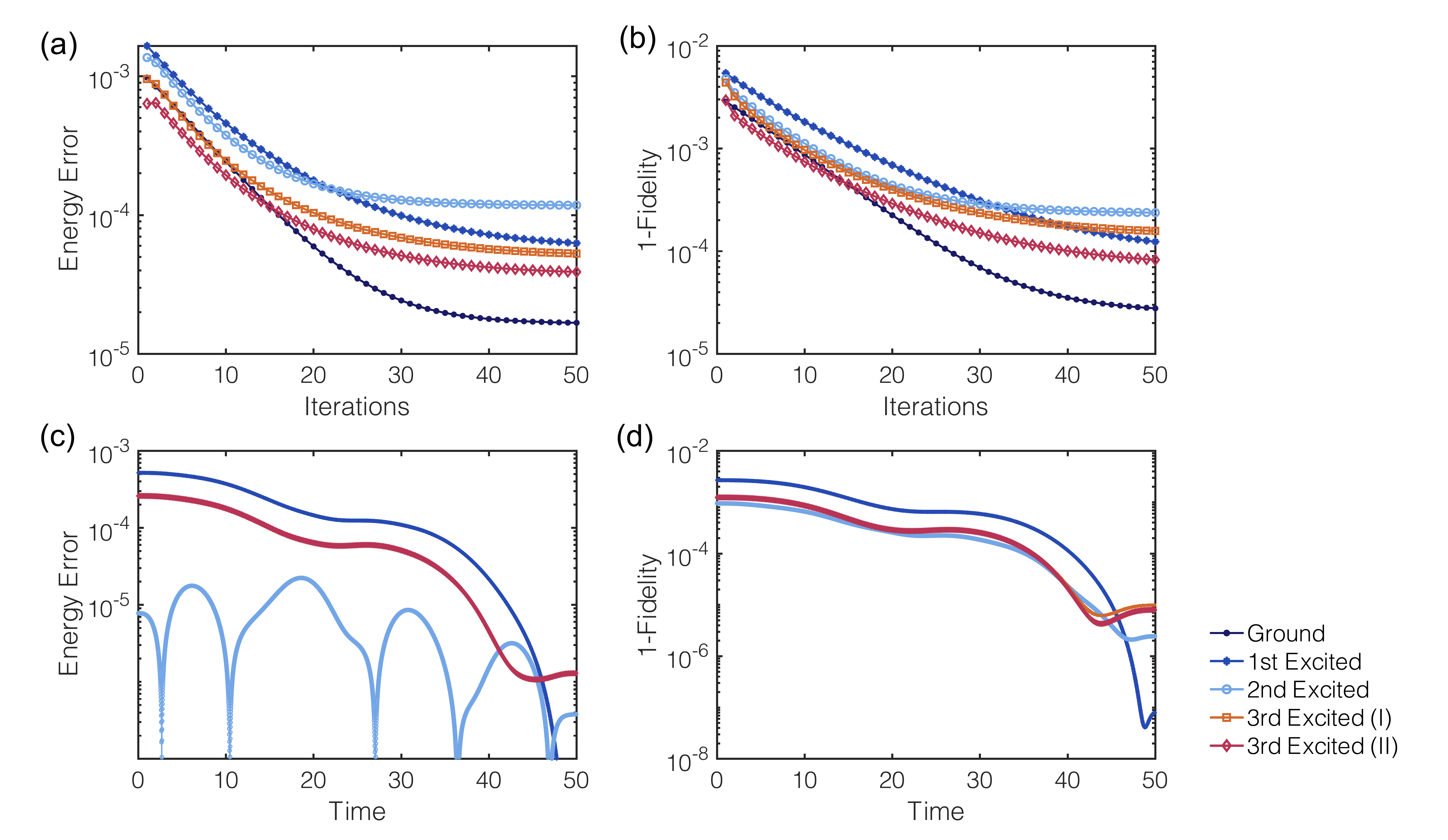}
\caption{ 
The ground state and excited states preparation of $\lambda \phi^4$ theory in the momentum space. The static mass is $m_0 = 0.5$ and the interacting strength is $\lambda_0 = 1$. 
The initial state for the excited states searching is the corresponding low-lying excited states of the free Hamiltonian $\lambda_0 = 0$.
(a,b) The convergence towards the ground state and excited states of $\lambda \phi^4$ theory in the momentum space. (a) The energy error towards iteration. (b) The fidelity error towards iteration. 
(c,d) The energy error (c) and the fidelity error (d) under adiabatic evolution from the initial state. 
}
\label{fig:mom_m0p5_1}
\end{figure}

Let us first consider $m_0 = 0.5$. In this regime, the first five  excited states of the free Hamiltonian are 
\begin{equation}\label{InitialState2}
\ket{1,0,0,0}, \ket{2,0,0,0}, \ket{0,1,0,0}, \ket{0,0,0,1}, \ket{3,0,0,0}.
\end{equation} 
Compared to the case of $m_0 = 1.27$, the two-particle states have lower energies than the single-particle state, and the state $\ket{0,0,1,0}$ is indeed a highly excited state (higher than three-particle states). 
The excited states of the interacting Hamiltonian with $\lambda_0 = 1$ follows the same order as that of the free Hamiltonian. We compare the simulation results for $\lambda_0 = 1$ using the variational methods and adiabatic evolution   in Figure~\ref{fig:mom_m0p5_1}.

 \begin{figure}[htbp]
\centering
\includegraphics[width=1.04\linewidth]{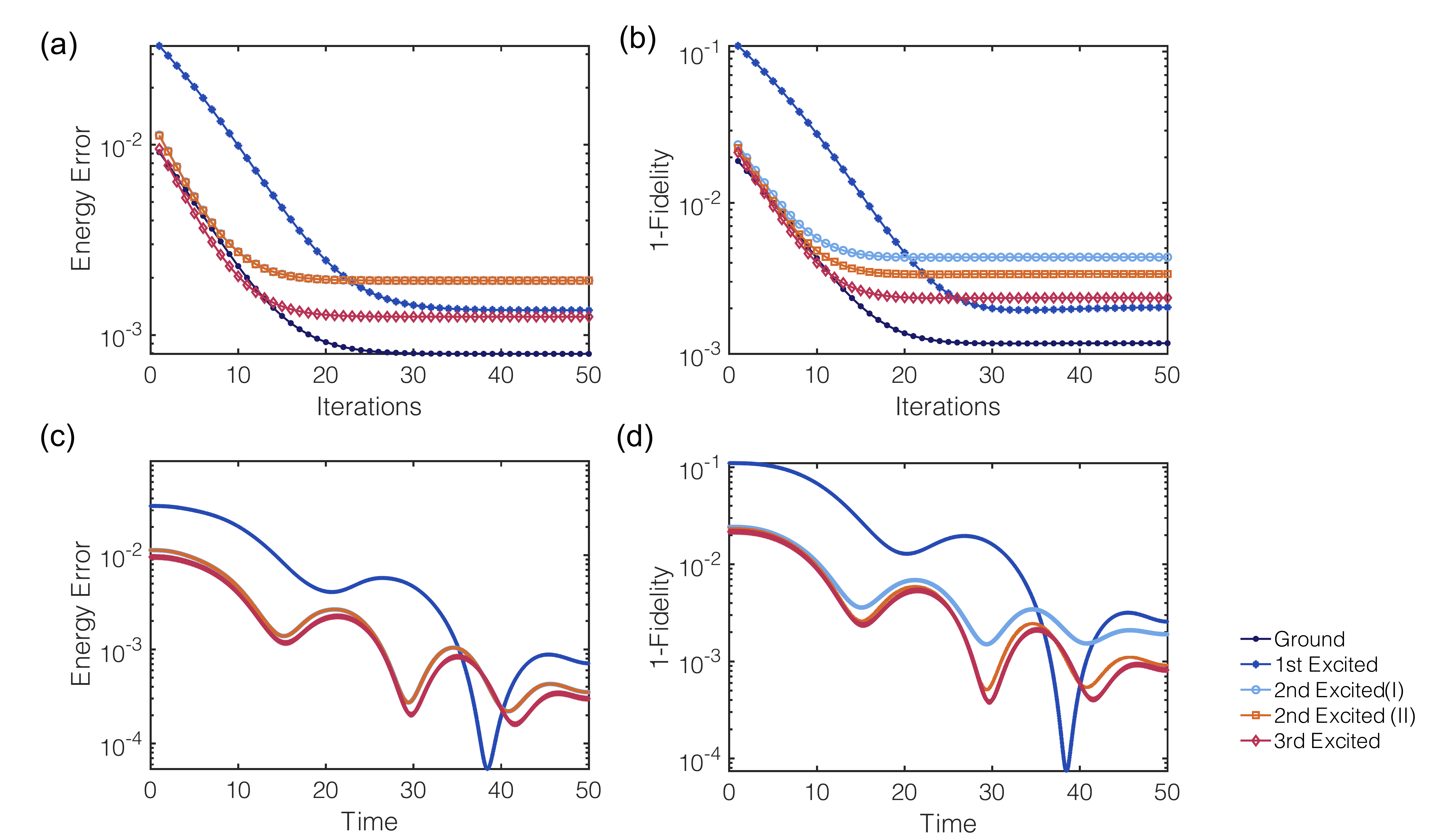}
\caption{ 
The static mass is $m_0 = 0.5$ and the interacting strength is $\lambda_0 = 10$. 
The initial state for the excited states searching is the single-particle excited states of the free Hamiltonian $\lambda_0 = 0$.
(a) The energy error towards iteration. (b) The fidelity error towards iteration. 
(c,d) The energy error (c) and the fidelity error (d) under adiabatic evolution from the initial state. 
}
\label{fig:mom_m0p5_10}
\end{figure}

Figure~\ref{fig:mom_m0p5_1}, (a,b) and (c,d) shows the results using variational methods and adiabatic evolution, respectively. 
Overall, the low-lying eigenstates obtained from the variational algorithms can be found with a high state fidelity.

In the strongly coupling regime, the two-particle excitation  $\ket{2,0,0,0}$ for the interacting Hamiltonian with $\lambda_0 = 10$ has higher energies compared to all the single-particle excitation, and it is even higher than the three-particle excitation $\ket{3,0,0,0}$. Therefore, we choose to prepare the single-particle excited states in the free theory as the initial states,
\begin{equation}\label{InitialState1}
\ket{1,0,0,0},  \ket{0,1,0,0}, \ket{0,0,0,1}, \ket{0,0,1,0}.
\end{equation}
Then, we use the variational quantum algorithms to search for the low-lying excited states. We show the simulation results with variational methods and adiabatic evolution for  $\lambda_0 = 10$ in Figure~\ref{fig:mom_m0p5_10}.
Figure~\ref{fig:mom_m0p5_10} shows that we could still use the excited state of the free Hamiltonian as the initial guess and obtain the target state with relatively high fidelity.

In the interacting theory, the single-particle excitation may not be well defined, and the eigenstate of the free theory may not be adequate for finding the ground state in the interacting theory, especially for the large interacting field $\lambda_0 = 10$. Note that the problem for the choice of the initial state exists in the adiabatic evolution.

\begin{figure}[htbp]
\centering
\includegraphics[width=1.04\linewidth]{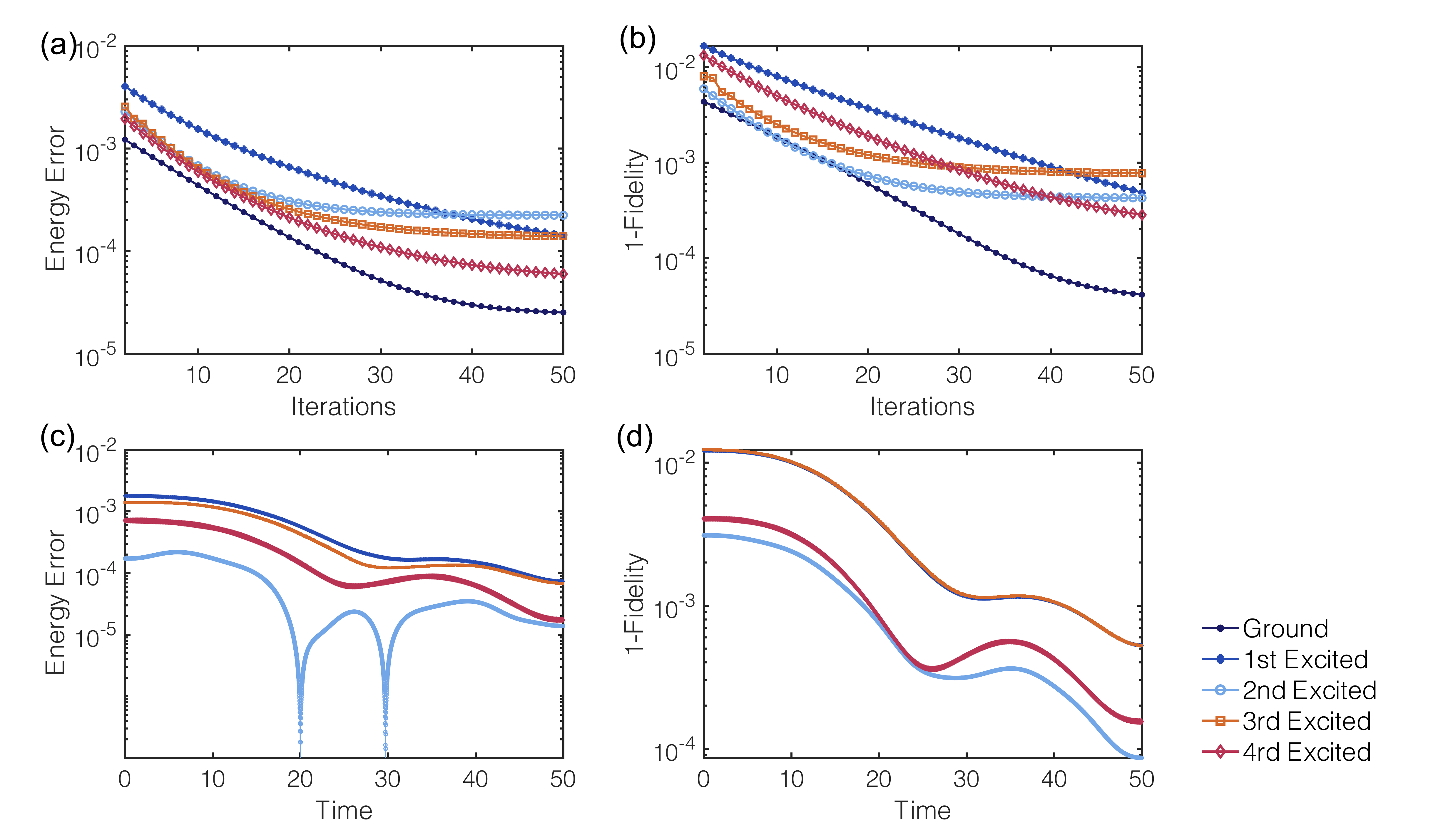}
\caption{ 
The static mass is $m_0 = 0.37$ and the interacting strength is $\lambda_0 = 1$. 
The initial state for the excited states searching is the corresponding low-lying excited states of the free Hamiltonian $\lambda_0 = 0$.
(a,b) The convergence towards the ground state and excited states. (a) The energy error towards iteration. (b) The fidelity error towards iteration. 
(c,d) The energy error (c) and the fidelity error (d) under adiabatic evolution from the initial state. 
}
\label{fig:mom_m0p37_1}
\end{figure}

Similarly, we could consider $m_0 = 0.37$, with the first five excited states of the free Hamiltonian as 
\begin{equation}
   \ket{1,0,0,0}, \ket{2,0,0,0}, \ket{3,0,0,0}, \ket{0,1,0,0}, \ket{0,0,0,1}.
\end{equation}
We first show the convergence towards iteration in Figure~\ref{fig:mom_m0p37_1} for $\lambda_0 = 1$ for both the variational state preparation and adiabatic state preparation. 

Similar to the case of $m_0 = 0.5$, the two-particle excitation for $m_0 = 0.37$ has a small energy in the free theory, but it has a much higher energy compared to all the single-particle excitations and even higher than the three-particle excitation in the strong coupling regime.
For instance, for a large interaction strength $\lambda_0 = 10$,  the energy has the following relation
\begin{equation}
    E\left(\ket{3,0,0,0}\right) < E\left(\ket{0,0,1,0}\right) < E\left( \ket{2,0,0,0}\right),
\end{equation}
which indicates the energy single-particle excitation is between the multi-particle excitation in the interacting field.
However, in the interacting field, we can find that the single-particle state is actually much close to the excited states in terms of state fidelity. 
Therefore, we similarly choose the initial states as the single-particle excited states.

 \begin{figure}[htbp]
\centering
\includegraphics[width=1.04\linewidth]{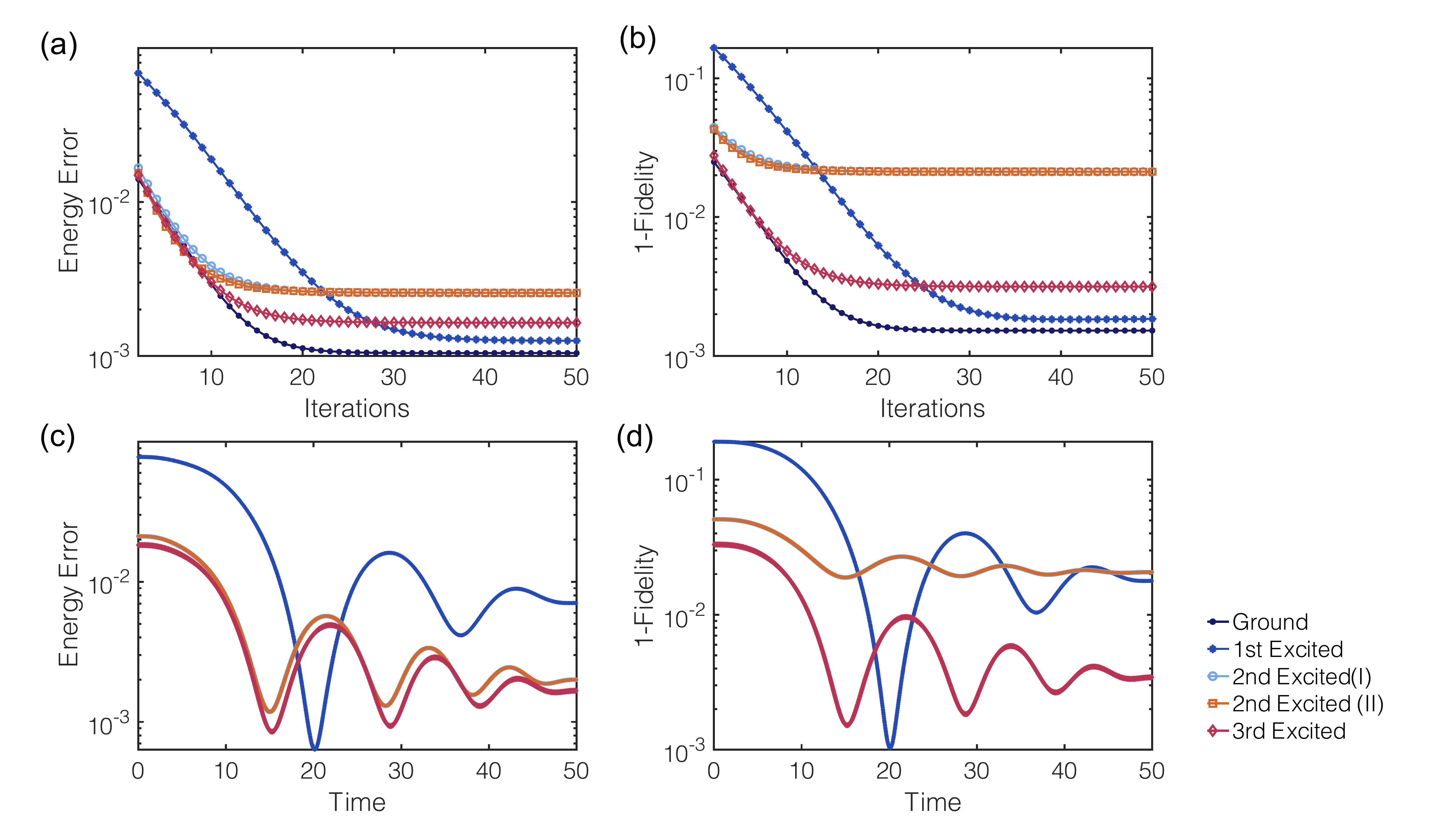}
\caption{ 
The static mass is $m_0 = 0.37$ and the interacting strength is $\lambda_0 = 10$. 
The initial state for the excited states searching is the corresponding single-particle excited states of the free Hamiltonian $\lambda_0 = 0$.
(a) The energy error towards iteration. (b) The fidelity error towards iteration. 
(c,d) The energy error (c) and the fidelity error (d) under adiabatic evolution from the initial state. 
}
\label{fig:mom_m0p37_10}
\end{figure}

 \begin{figure}[htbp]
\centering
\includegraphics[width=0.86\linewidth]{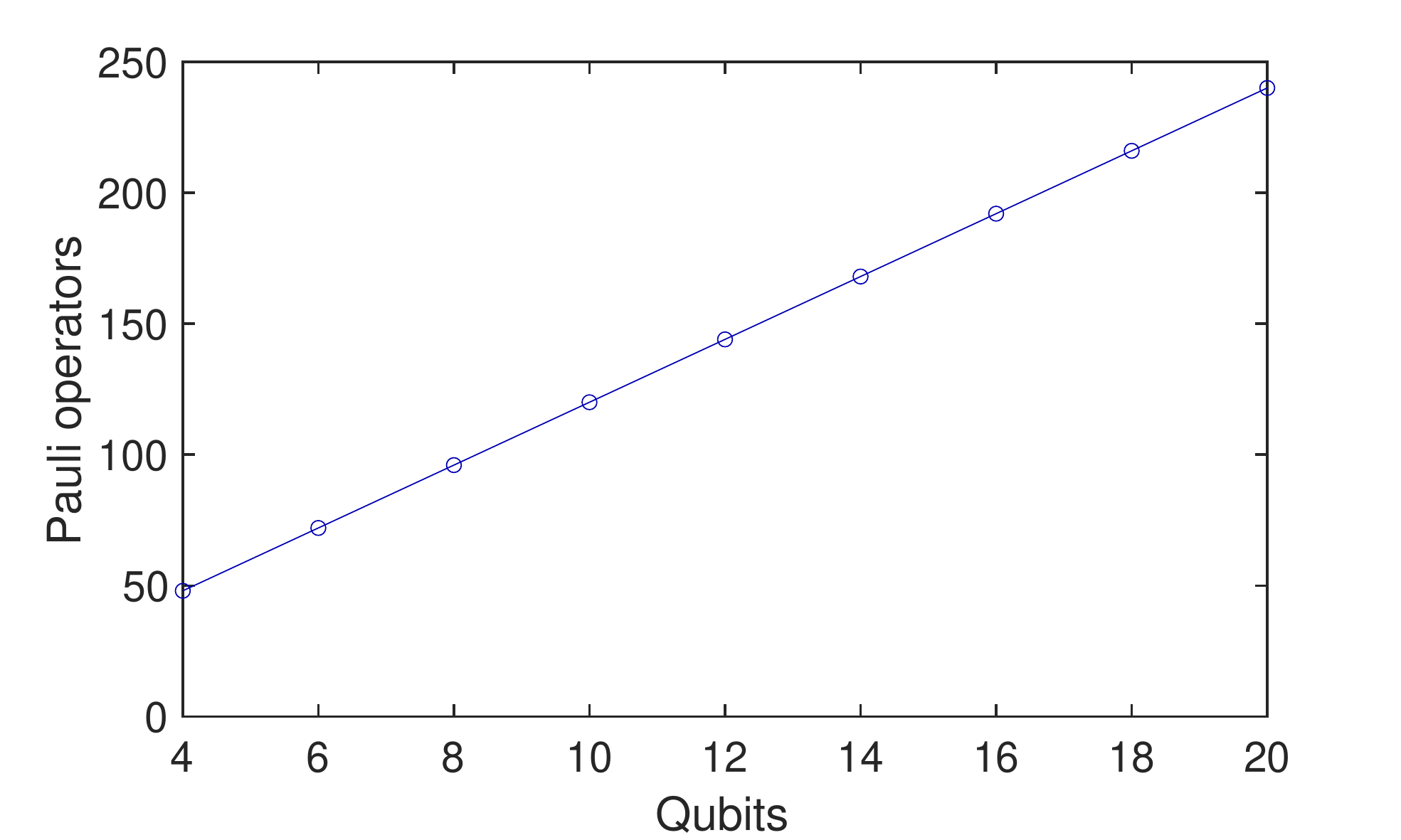}
\caption{The number of Pauli operators in the variational circuits.
}
\label{fig:pauli_term}
\end{figure}

We show the convergence towards iteration in Figure~\ref{fig:mom_m0p37_10} for $\lambda_0 = 10$. 
As shown in Figure~\ref{fig:mom_m0p37_10} (b,d), we can find that the state fidelity of the second excited state and the third excited state are relatively lower than the others. This is because these two states are actually evolved from the two degenerate states due to the boundary condition in the momentum space. However, the state fidelity in the subspace spanned by the degenerate states is numerically tested to be over $99\%$.

\sun{ 
	Finally, we show a detailed resource analysis of our method. In the simulation, we can reduce the number of Pauli operators by restrict the higher order transitions. Here, we consider to fix the energy cutoff as $n_{\mathrm{cut}} = 4$ and we restrict the higher order transition to be less than $3$ in $\hat T_1$ and the modified $\hat T_2$ operator. Then the energy constraint $\vert s_i - t_i \vert \leq 4$ is trivially hold and thus we need $6$ Pauli operators to construct a single term from $\hat T_1$ (they are $\sigma_x \sigma_x, \sigma_y \sigma_y, \sigma_x \sigma_z, \sigma_x I, \sigma_z \sigma_x, I \sigma_x$ ). A linear combinations of tensor products of all pairs of these Pauli operators to form a single term from Eq.~\eqref{eq:UVCC_pair}. In practice, we may only use the second term in Eq.~\eqref{eq:UVCC_pair}, which saves half resources. Therefore, the number of Pauli rotation operators required in $\hat T_2$ is upper bounded by $18 N$. Figure.~\ref{fig:pauli_term} shows the number of Pauli rotation operators in the $N$-qubit variational circuit $\hat T_1 + \hat T_2$.
}

\section{Outlooks}\label{outlook}
In this paper, we discuss constructions of a variational version of the Jordan-Lee-Preskill algorithm in the near-term quantum computer. We justify the validity of the algorithm by several numerical simulations. We believe that our hybrid quantum-classical algorithm will eventually benefit possible solutions to open problems in fundamental physics, and benchmark tasks of near-term quantum devices. Here, we summarize some potential research directions along our path. 

\subsection{Relation to the physical observables}
In the previous discussions, we demonstrate the numerical simulation with fixed lattice spacing and lattice sites. To obtain the expectation value of physical observables in the real scalar field, say $\left\langle \phi  \right\rangle$, we can first measure the expectation value with a series of increasing number of sites $N$, $\left\langle {\phi _{a,N}^{(0)}} \right\rangle $, and  extrapolate to the infinite volume limit $N \rightarrow \infty$, $\left\langle {\phi _{a,N \to \infty }^{(0)}} \right\rangle $. Then, we further extrapolate these results to the continuum limit $a \rightarrow 0$, $\left\langle {{\phi ^{(0)}}} \right\rangle  \equiv \left\langle {\phi _{a \to 0,N \to \infty }^{(0)}} \right\rangle $. Finally, we renormalize the expectation value as $\left\langle {{\phi ^{(R)}}} \right\rangle  = Z\left\langle {{\phi ^{(0)}}} \right\rangle $ with the renormalization constant $Z$.
We leave the discussions to future work.

\subsection{Simulating quantum field theories in the NISQ era}
Our work opens up a new direction of formulating several quantum field theory tasks in the setup of variational quantum simulation. In the era of noisy intermediate-scale quantum (NISQ), we expect that hybrid, variational quantum simulation algorithms might be one of the most accessible ways regarding near-term quantum hardware. 

There is a landscape of quantum field theories with a full basket of open problems, who are looking for the potential computational capacity of quantum devices. Our work about strongly-coupled $\lambda \phi^4$ theory is one of the simplest examples, whose non-perturbative nature is not fully understood by quantum field theorists. One could consider generalizing the scattering paradigm and its relevant techniques towards other quantum field theories. Specifically, lattice gauge theories in the four dimensions are particularly important for particle physicists, since it is related to quantum chromodynamics (QCD) and the Standard Model in particle physics.
We refer to Refs.~\cite{shaw2020quantum,kokail2019self,farrelly2020discretizing,lamm2019general,chakraborty2020digital,hauke2013quantum,Paulson_2021,surace2020lattice} for recent theoretical and experiment advances in the quantum simulation of lattice gauge field theories.
One could look for other strongly-coupled quantum field theory problems, for instance, phase transitions in the finite-temperature quantum field theories that are closely related to nuclear physics \cite{Preskill:2018fag,Alexeev:2019enj}.

\subsection{Identifying possible quantum advantages}
Variational algorithms running on near-term devices might have further advantages for fundamental studies in quantum information science. Specifically, since we could design hybrid quantum-classical algorithms, it is easy for us to diagnose which classical or quantum steps have advantages practically. Although in this work, we do not focus on this comparison, we expect that similar comparisons could be performed in future studies. In the future, people might work out practically, which steps in the whole algorithms might have the quantum advantage, and if so, how much advantage they will have. Those studies might be helpful to construct the most useful quantum algorithms using practical experiences, and use those experiences to benchmark near-term devices. Quantum simulation of quantum field theories is a field that is still young, but we expect that finally, more techniques and hardcore developments could be formulated (see some similar analysis in computational quantum chemistry \cite{von2020quantum}).

\subsection{Error mitigation}

In this paper, we neglected errors from device imperfections and shot noise from finite measurement samples. Those errors could accumulate and affect the simulation accuracy. Fortunately, various error mitigation techniques have been developed to suppress device errors \cite{subspace1,Li2017,bonet2018low,PhysRevLett.119.180509, endo2017practical, subspace2,recoveringnoisefree,PhysRevA.99.012334, samerrormitigation,bonet2018low,sun2020mitigating2,sugurureview21, cerezo_variational_2021}. By properly post-processing measurement results from different circuit realizations (e.g., with different noise ratios or symmetries), one can suppress the effect of noise by several orders~\cite{kandala2019error,1084,kim2021scalable,bonet2018low}.  {For instance, the $\lambda \phi^4$ field preserves the reflection symmetry, so one can project the quantum state in the symmetry-protected subspace (see, e.g.,~\cite{subspace1,bonet2018low,sugurureview21})}

Meanwhile, the effect of shot noise could be reduced as well by exploiting more advanced measurement schemes \cite{kandala2017hardware,wu2021overlapped,verteletskyi2020measurement,hadfield2020measurements,torlai2020precise,huang2020predicting,huang2021efficient,cotler2020quantum,hadfield2021adaptive, hillmich2021decision,zhang2021experimental}. The basic idea is to either exploit observable compatibility, importance sampling, or additional quantum circuit to more efficiently measure observables. Combining those error mitigation and advanced measurement schemes, we might be able to demonstrate our algorithms with current or near-term quantum hardware.\\

\section{Acknowledgements}
{Acknowledgements.}---This paper is mostly finished when JL is a graduate student in Caltech. We thank Alex J. Buser, Liang Jiang, Natalie Klco, Peter Love, Ash Milsted, John Preskill, Burak Sahinoglu, Guifre Vidal, and Xiaoyang Wang for related discussions. JL is supported in part by the Institute for Quantum Information and Matter (IQIM), an NSF Physics Frontiers Center (NSF Grant PHY-1125565) with support from the Gordon and Betty Moore Foundation (GBMF-2644), the Walter Burke Institute for Theoretical Physics. JL is also supported in part by International Business Machines (IBM) Quantum through the Chicago Quantum Exchange. XY acknowledges support from the Simons Foundation. \\

\noindent{Note added.}---Around the time when this research is finished, the papers \cite{Macridin:2021uwn,kurkcuoglu2021quantum} appear, which has some overlaps with discussions in our paper, including the field bases~\cite{Macridin:2021uwn} and quantum simulation using qudits~\cite{kurkcuoglu2021quantum}.

\bibliographystyle{apsrev4-1}
\bibliography{Biblio.bib}

\pagebreak
\clearpage
\foreach \x in {1,...,\the\pdflastximagepages}
{
	\clearpage
	

\section*{Supplementary material}

# Appendix

(Dated: June 29, 2022)

## CONTENTS

## I. LSZ REDUCTION FORMULA

We briefly describe the use of Harmonic oscillator basis in calculating scattering amplitudes in the scalar field theory by *Lehmann-Symanzik-Zimmermann reduction formula*. We start from an important observation: consider the two-point function: $\langle \Omega | \phi(x)\phi(y) | \Omega \rangle$, where $|\Omega\rangle$ is the ground state or the physical vacuum. From a non-perturbative perspective, computing the two-point function will tell us the information about the masses of particles through the *Källén-Lehmann spectral representation* [**? ?**], from which we can find and compute the *field strength* renormalization factor. Let us consider a complete basis describing the spectrum of the interacting $\phi^4$ theory by the spectrum $|n_{\mathbf{p}}\rangle$ indexed by the d-momentum $\mathbf{p}$ (recall Lorentz invariance so that $|n_{\mathbf{p}}\rangle$ is the simultaneous eigenstate for both $H$ and $d$-momentum operator $\mathbf{P}$):

$$\hat{H} |n_{\mathbf{p}}\rangle = \omega_{\mathbf{p}}(n) |n_{\mathbf{p}}\rangle,$$
$$\hat{P} |n_{\mathbf{p}}\rangle = \mathbf{p} |n_{\mathbf{p}}\rangle, \tag{I.1}$$

where the spectrum eigenstates can be: *(a)* single particle state with a renormalized mass $m$ and $\omega_p = \sqrt{m^2 + p^2}$, *(b)* bound states, and *(c)* multi-particle states forming from single-particle or bound states. All these states $|n_{\mathbf{p}}\rangle$ are excited from the ground state $|\Omega\rangle$, which can be evolved from the ground state of the free theory:

$$\lim_{t_- \to -\infty} \exp(iHt_-) |0\rangle = \lim_{t_- \to -\infty} |\Omega\rangle \langle\Omega|0\rangle + \sum_{n_{\mathbf{p}}} |n_{\mathbf{p}}\rangle \langle n_{\mathbf{p}}|0\rangle \exp^{-i\omega_p t_-}$$
$$= |\Omega\rangle \langle\Omega|0\rangle, \tag{I.2}$$

where the second equation is obtained from the Riemann–Lebesgue lemma. Thus we can compute the two-point correlator at interaction by:

$$\langle\Omega|\phi(x)\mathbb{I}\phi(y)|\Omega\rangle = \sum_n \int \frac{d^{d+1}\mathbf{p}}{(2\pi)^{d+1}} \frac{1}{2\omega_{\mathbf{p}}(\lambda)} \langle\Omega|\phi(x)|n_{\mathbf{p}}\rangle \langle n_{\mathbf{p}}|\phi(y)|\Omega\rangle, \tag{I.3}$$

where we normalize the field: $\phi(x) \mapsto \phi(x) - \langle\Omega| \phi(x) |\Omega\rangle$ and:

$$\mathbb{I} = |\Omega\rangle \langle\Omega| + \sum_n |n_{\mathbf{p}}\rangle \langle n_{\mathbf{p}}|. \tag{I.4}$$

The identity operator constructed from the interaction spectrum. Note that we have:

$$\langle\Omega| \phi(x) |n_{\mathbf{p}}\rangle = \langle\Omega| \exp(i\hat{\mathbf{P}}x)\phi(0)\exp(-i\hat{\mathbf{P}}x) |n_{\mathbf{p}}\rangle = \langle\Omega| \phi(0) |n_{\mathbf{p}}\rangle \exp(-i\omega_{\mathbf{p}}(n)x). \tag{I.5}$$

To work in the rest frame, we consider the Lorentz boost: $|n_0\rangle = U |n_{\mathbf{p}}\rangle$ so that we can further rewrite the RHS of Eq.(I.3):

$$\langle\Omega| \phi(x) |n_{\mathbf{p}}\rangle = \langle\Omega| U^\dagger U \phi(0)U^\dagger U |n_{\mathbf{p}}\rangle \exp(-i\mathbf{p}x)$$
$$= \langle\Omega| \phi(0) |n_0\rangle \exp(-i\mathbf{p}x), \tag{I.6}$$

so that we can express the two-point function with time-ordered a free part and a spectral function $\rho(M^2)$:

$$\langle\Omega|\,\mathcal{T}\phi(x)\phi(y)\,|\Omega\rangle = \int\frac{dM^2}{2\pi}\rho(M^2)\int\frac{d^{d+1}p}{(2\pi)^{d+1}}\frac{1}{2\omega_{\mathbf{p}}(n)}e^{-ip(x-y)},\tag{I.7}$$

where now we integrate over $d+1$ energy-momentum and the spectral function defined:

$$\rho\left(M^2\right)=\sum_n 2\pi\delta\left(M^2-m_n^2\right)|\langle\Omega|\phi(0)|n_0\rangle|^2\,.\tag{I.8}$$

The spectrum function leads to isolated peak of Driac Delta functions for the single-particle state where $M^2$ is less than $4m^2$ or $m_{\mathrm{bound}}^2$ at the rest frame $|\mathbf{p}_0\rangle$:

$$\rho\left(M^2\right)=2\pi\delta\left(M^2-m^2\right)|\langle\Omega|\phi(0)|\mathbf{p}_0\rangle|^2\,.\tag{I.9}$$

The *field-strength* renormalization is given by:

$$Z_3\equiv|\langle\Omega|\phi(0)|\mathbf{p}_0\rangle|^2=|\langle\Omega|\phi(0)|\mathbf{p}\rangle|^2\,,\tag{I.10}$$

where the last equality follows the fact that the spectrum function is Lorentz invariant. One can prove that $Z_3\le 1$ and equals one if and only if it is in the free theory. As a result, to even write down the two-point function, we require the field to be renormalized:

$$\phi(x)\to\frac{1}{\sqrt{Z_3}}(\phi(x)-\langle\Omega|\,\phi(x)\,|\Omega\rangle).\tag{I.11}$$

In the latter part, we always assume $\phi(x)$ is properly renormalized. In simulating scattering amplitude of the $\phi^4$ theory in quantum circuits, one should, therefore, always first measure the two-point function in term of the self-energy to obtain the proper field-strength renormalization factor [**?** ]. This is an important caveat to measuring the scattering amplitude in quantum circuits using the Harmonic oscillator basis. In scattering theory, the use of harmonic oscillator basis is to prepare in-and-out states (particularly in-states). It is conventional to assume that the in-states start far past as a wave packet and then evolve in time as free fields to interaction, as a consequence of the locality of interaction. Therefore, the use of harmonic oscillator in quantum circuits is to encode the far past in-states, where we now use the Heisenberg picture to denote the fields $\phi_{\mathrm{in}}^H(t,x)$ ($\phi_{\mathrm{out}}^H(t,x)$). Then we can express the creation and annihilation operators at some arbitrary time instance $t_0$:

$$\begin{aligned}a_{\mathbf{p}}(t_0)&=\int d^{d+1}xe^{ipx}[i\partial_0\phi^H+\omega_{\mathbf{p}}\phi^H]|_{t=t_0},\\a_{\mathbf{p}}^\dagger(t_0)&=i\int d^{d+1}xe^{-ipx}[-\partial_0\phi^H+\omega_{\mathbf{p}}\phi^H]|_{t=t_0}.\end{aligned}\tag{I.12}$$

Note that at $t_0\to-\infty$, the above equation simply implies the mode expansion of the scalar fields in the free theory where we can express the scattering amplitudes:

$$\langle q_1,\cdots,q_m;\mathrm{out}\,|p_1,p_2,\cdots,p_n,\mathrm{in}\rangle=\langle\Omega|\,a_{\mathbf{q}_1}(t_{\mathrm{f}})\cdots a_{\mathbf{q}_m}(t_{\mathrm{f}})a_{\mathbf{p}_1}^\dagger(t_{\mathrm{i}})\cdots a_{\mathbf{p}_n}^\dagger(t_{\mathrm{i}})\,|\Omega\rangle\,.\tag{I.13}$$

The variational ansätze constructed in this paper can simulate initial states $|i\rangle$ specified momentum sector as well as a wave-packet:

$$|i;\mathrm{in}\rangle=\int\frac{d^d\mathbf{k}}{(2\pi)^d}\frac{1}{2w_k}F(\mathbf{k})\,|k\rangle\to\sum_{k\in\Gamma}\frac{1}{L}\frac{1}{2\omega_k}F(k)\,|k\rangle\,.\tag{I.14}$$

The specific form of the wave-packets should not concern us here; however, it is an independent issue to analyze the encoding of $F(k)$ in the quantum circuits if we were to prepare a wave-packets in the Harmonic oscillator basis. We cam analogously define the wave-packet creation operator:

$$\int d^d\mathbf{k}F(\mathbf{k})a_{\mathbf{k}}^\dagger(t)\,|\Omega\rangle\equiv a_F^\dagger(t)\,|\Omega\rangle\,.\tag{I.15}$$

Notice that the wave-packets must preserve the inner products; thus, in a quantum circuits with harmonic oscillator basis, one can write:

$$Va_{\mathbf{k}}^\dagger(t)V^\dagger=a_F^\dagger(t)\tag{I.16}$$

For some unitary matrix $V$ also defined on the harmonic oscillator basis. Note that:

$$a_F^\dagger(\infty) - a_F^\dagger(-\infty) = \int_\infty^\infty dt \partial_0 a_F^\dagger(t),$$ (I.17)

where the LHS can be directly applied to Eqn.(I.13) where $\langle \Omega | \, a^\dagger(\infty) = 0$ and similarly $a(-\infty) \, |\Omega\rangle = 0$. The RHS can be evaluated from the wave-packets Eqn.(I.15) and Eqn.(I.12):

$$
\begin{aligned}
RHS &= \int d^d\mathbf{k} F(\mathbf{k}) \int_{-\infty}^\infty dt \partial_0 \alpha_F^\dagger(t) = \int d^d\mathbf{k} F(\mathbf{k}) \int d^{d+1}x \partial_0 e^{-ikx}[-i\partial_0 + \omega_\mathbf{k}]\phi(x) \\
&= \int d^d\mathbf{k} F(\mathbf{k}) \int d^{d+1}x \left\{ (-i\omega_\mathbf{k})e^{-ikx}[-i\partial_0 + \omega_\mathbf{k}]\phi + e^{-ikx}[-i\partial_0^2 + \omega_\mathbf{k}\partial_0]\phi \right\} \\
&= i\int d^d\mathbf{k} F(\mathbf{k}) \int d^{d+1}x e^{-ikx}(\partial^2 + m^2)\phi,
\end{aligned}
$$ (I.18)

where we use the convention $dtdx = d^2x$ and the integration by parts in:

$$\int d^{d+1}x(-i\omega_k)e^{-ikx}(-i\partial_0\phi) = i\int d^{d+1}x e^{-ikx}(\partial_0^2\phi).$$ (I.19)

Note that if we are describing a free (on-shell) theory, $(\partial^2 + m_0^2)\phi = 0$ implies there is no quantization, i.e., no particles being created or annihilated in the process. In the interaction, the on-shell condition is replaced by some external source $j(x)$ such that:

$$(\partial^2 + m^2)\phi(x) = j(x).$$ (I.20)

Analogously, let assume that the final state is described by some wave-packets:

$$|f\rangle \equiv \lim_{t\to\infty} \int d^d\mathbf{k} G(\mathbf{k}) a_\mathbf{k}^\dagger(t) \, |\Omega\rangle \equiv \lim_{t\to\infty} a_G^\dagger(t) \, |\omega\rangle \, .$$ (I.21)

We would can derive the LSZ formula by putting everything together:

$$
\begin{aligned}
\langle f|i\rangle &= \lim_{\substack{t_f\to\infty \\ t_i\to-\infty}} \langle \Omega | \, a_G(t_f) a_F^\dagger(t) \, |\Omega\rangle \\
&= i^2 \int d^d\mathbf{q} d^d\mathbf{p} G(\mathbf{q}) F(\mathbf{p}) \int d^{d+1}y d^{d+1}x \left( e^{iqy}e^{-ipx} \right) (\partial_y^2 + m^2)(\partial_x^2 + m^2) \langle \Omega | \, \mathcal{T}\phi^H(t_f, y)\phi^H(t_i, x) \, |\Omega\rangle \, .
\end{aligned}
$$ (I.22)

Therefore, we can express the more general version of the scattering amplitudes from Eq.(I.13):

$$
\begin{aligned}
&\langle q_1, \cdots, q_m; \text{out} \, |p_1, p_2, \cdots, p_n, \text{in}\rangle = (i)^{m+n} \int d^{d+1}y_1 \cdots d^{d+1}y_m d^{d+1}x_1 \cdots d^{d+1}x_n \left( e^{iq_1y_1} \cdots e^{iq_my_m} e^{-ip_1x_1} e^{-ip_nx_n} \right) \\
&(\partial_{y_1}^2 + m^2) \cdots (\partial_{y_m}^2 + m^2)(\partial_{x_1}^2 + m^2) \cdots (\partial_{x_n}^2 + m^2) \langle \Omega | \, \mathcal{T}\phi(y_1) \cdots \phi(y_m)\phi(x_1) \cdots \phi(x_n) \, |\Omega\rangle \\
&= (i)^{m+n}(q_1^2 + m^2) \cdots (q_m^2 + m^2)(p_1^2 + m^2) \cdots (p_n^2 + m^2) \langle \Omega | \, \mathcal{T}\tilde{\phi}(q_1) \cdots \tilde{\phi}(q_m)\tilde{\phi}(p_1) \cdots \tilde{\phi}(p_n) \, |\Omega\rangle \, ,
\end{aligned}
$$ (I.23)

where at the last step we assume the wave-packets to be the planar wave functions localized at $\delta^d(\mathbf{q}_1) \cdots \delta^d(\mathbf{q}_m)\delta^D(\mathbf{p}_1) \cdots \delta^D(\mathbf{p}_n)$. The $m + n$-point function can be calculated from the partition function perturbatively from Feynman Diagrams. Combing with that, we can derive rigorously the Feynman rules. In particular, the LSZ reduction formula implies that it, to obtain the scattering amplitude, is sufficient to apply using Eq. (I.2):

$$
\begin{aligned}
\lim_{t_0\to-\infty}[U(t,t_0)a_p(t_0)U(t,t_0)]^\dagger \, |\Omega\rangle &= \lim_{t_0\to-\infty} U(t,t_0)a_p^\dagger(t_0) \, |0\rangle \, (\langle\Omega|0\rangle)^{-1} \\
&= U(t)/(\langle\Omega|0\rangle)^{-1} \, |p; \text{free}\rangle \, .
\end{aligned}
$$ (I.24)

Simulating the above evolution from free states would ensure to prepare the proper in states for the scattering amplitudes, where the factor $(\langle\Omega|0\rangle)^{-1}$ normalizes the vacuum bubbles.

The UCC ansätze used in this work can be seen to create a wave-packet in interaction region with some fixed $t$. Recall Eq.(I.16), we can always decompose our UCC ans"atze into the following form (See in Eq.(24) & Eq.(25) & Eq.(26) in main text)

$$\exp(\hat{T}_\ell(\theta(t)) - \hat{T}_\ell^\dagger(\theta(t))) = \lim_{t_0\to-\infty} VU(t,t_0)$$ (I.25)

Note that $\hat{T}_4(\theta(t))$ in the adiabatic picture can be used to simulate the full Hamiltonian (Eq.(21) in the main text); hence, in which case $V$ is the identity operator. Therefore, Eq.(I.24) provides a simple interpretation of the use of variational UCC ansätze: to create a wave-packet at some intermediate time $t$, for which we can denote as *initial time* $t_i$ of the optimization process using variational quantum imaginary time evolution [?] or other optimization method such as [?]. The optimization would produce a final time $t_f$ (or $\tau_f \equiv it_f$), where we further evolve the wave-packet into lowest energy eigenstates where one can utilize physic prior to recognizing these states and compute corresponding scattering amplitudes.

## II. TRUNCATION ERROR OF FIELD BASIS

We provide a rigorous analysis on the truncation error of free theory eigenstates under field basis and compare with the case of harmonic oscillator basis. Let $|\psi\rangle$ be any state expanded under field basis as

$$|\psi\rangle = \int_{-\infty}^{\infty} d\phi_1 \cdots \int_{-\infty}^{\infty} d\phi_N \psi\left(\phi_1, \ldots, \phi_N\right) |\phi_1, \ldots, \phi_N\rangle,$$

with truncated approximation

$$|\psi_{\text{cut}}\rangle = \int_{-\phi_{\max}}^{\phi_{\max}} d\phi_1 \cdots \int_{-\phi_{\max}}^{\phi_{\max}} d\phi_N \psi\left(\phi_1, \ldots, \phi_N\right) |\phi_1, \ldots, \phi_N\rangle.$$

Then the truncation error $\epsilon$ can be defined as

$$1 - \epsilon = \langle\psi|\psi_{\text{cut}}\rangle = \int_{-\phi_{\max}}^{\phi_{\max}} d\phi_1 \cdots \int_{-\phi_{\max}}^{\phi_{\max}} d\phi_N |\psi\left(\phi_1, \ldots, \phi_N\right)|^2 |\phi_1, \ldots, \phi_N\rangle.$$

For simplicity, we assume $N = 1$ which corresponds to the case of a free quantum harmonic oscillator. The eigenstates are given by Hermit-Gauss functions:

$$\psi_n(x) = \frac{1}{\pi^{\frac{1}{4}} \sqrt{2^n n!}} e^{-\frac{x^2}{2}} H_n(x),$$

where $H_n(x)$ are physicists' Hermit polynomials defined as

$$H_n(x) = (-1)^n e^{x^2} \frac{d^n}{dx^n} e^{-x^2}.$$

An explicitly evaluated integral of the p.d.f. $|\psi_n(x)|^2$ of the wavefunction $\psi_n(x)$ on a truncated domain can be found in [?]:

$$\epsilon(x, n) = 2 \int_x^{\infty} |\psi_n(y)|^2 dy = \text{erfc}(x) + \frac{1}{\sqrt{\pi}} e^{-x^2} \sum_{j=1}^n \binom{n}{j} \frac{H_{2j-1}(x)}{2^j j!}, \tag{II.1}$$

where

$$\text{erfc}(x) = 1 - \frac{2}{\sqrt{\pi}} \int_0^x e^{-y^2} dy$$

is the complementary error function. Our job is finding an upper bound of Eq. II.1. First of all, $\text{erfc}(x)$ is bounded by $\frac{2}{\sqrt{\pi}} e^{-x^2}$ for any $x \in \mathbb{R}$. To verify this point, let

$$f(x) = \frac{2}{\sqrt{\pi}} e^{-x^2} - \text{erfc}(x) \text{ with } f'(x) = \frac{2}{\sqrt{\pi}}(-2x + 1) e^{-x^2}.$$

The derivative is nonnegative in $[0, \frac{1}{2}]$ while negative in $(\frac{1}{2}, +\infty)$. Accordingly, $f$ is nonnegative in $(0, 1]$. We only need to make sure that $f$ will not turn into negative later. Note that by L'Hôpital's rule, $f(x) \to 0$ when $x \to +\infty$. If $f(x)$ could be negative, it must admit a global minimum at $x_0 \in (\frac{1}{2}, +\infty)$ and at which $f''(x_0) = 0$. However, this is impossible and hence the inequality holds on $\mathbb{R}$.

Now the problem is analyzing how would the remaining term containing $H_{2j-1}(x)$ decay with respect to both $x$ and $n$. As Eq. II.1 converts $H_n(x)^2$ from the integrand into $H_{2j-1}(x)$, it allows us to perform Cramer inequality [?]: $|\psi_{2j-1}(x)| \leq \pi^{-\frac{1}{4}}$ for all $x \in \mathbb{R}$. With this fact we have

$$\frac{1}{\sqrt{\pi}} e^{-x^2} \sum_{j=1}^n \binom{n}{j} \frac{H_{2j-1}(x)}{2^j j!} = \frac{1}{\pi^{\frac{1}{4}}} e^{-\frac{x^2}{2}} \sum_{j=1}^n \binom{n}{j} \frac{\sqrt{2^{2j-1}(2j-1)!}}{2^j j!} \psi_{2j-1} \leq \frac{1}{\sqrt{\pi}} e^{-\frac{x^2}{2}} \sum_{j=1}^n \binom{n}{j} \frac{\sqrt{2^{2j}(2j)!}}{2^j j!}.$$

Explicitly expansion shows that

$$\frac{\sqrt{(2j)!}}{j!} \leq \sqrt{2^{2j}} \implies \frac{\sqrt{2^{2j}(2j)!}}{2^j j!} \leq 2^{\frac{j}{2}} \implies \sum_{j=1}^n \binom{n}{j} \frac{\sqrt{2^{2j}(2j)!}}{2^j j!} \leq 2^{\frac{3n}{2}} \approx e^n.$$

Therefore,

$$\epsilon(x, n) \leq \frac{2}{\sqrt{\pi}} e^{-x^2} + \frac{1}{\sqrt{\pi}} e^{-\frac{x^2}{2}} e^n \approx C e^{-x^2 + Dn}. \tag{II.2}$$

For fixed energy level $n$, an exponentially small error can be obtained when $x > O(\sqrt{n})$. In the field context, this $x$ is though as $\phi_{\max}$ which is given as $2^{n_q}$ (multiplying with a chosen increment $\delta_\phi$) on a quantum computer with $n_q$ qubits. Thus we can conclude that $n_q = O(\log \log \frac{1}{\epsilon})$ of the error for approximating low-lying excited states of free theory and the number of qubits. As a comparison, the original truncation error bound of Jordan-Lee-Preskill algorithm is obtained by Chebyshev's inequality. As they make no presupposition on the concerned wavefunction, Chebyshev's inequality only yields a polynomial decay on the error.

The scaling $O(\sqrt{n})$ coincides with what indicates by WKB approximation: the classically forbidden region is given by the square root of energy. If one takes $x \approx O(\sqrt{n})$, the above inequality (II.2) should be further refined and numerical results from [? ] indicate that the error only decays polynomially for such cases.

When we change to harmonic oscillator basis, there is no such a truncation error on the number eigenstates because, as we explained in the main text, each of them is simply a single computational basis element $|s^{(k)}\rangle$ and there is no need to talk about truncating Hermit-Gauss functions in this setting. This is another advantage to work directly with harmonic oscillator basis. However, the error still stems from elsewhere. For instance, the creation/annihilation operators do not satisfy the commutation relation $[a^\dagger, a] = iI$ rigorously (this also happens for the field operator $\phi(x)$ and its conjugate momentum $\pi(x)$). Expanding the Lie bracket explicitly by matrices, one can check that this error $\epsilon'$ decays exponentially with the dimension of the local Hilbert space, and hence we have $n_q = \log \frac{1}{\epsilon'}$ where $n_q$ is the number of qubits per site.

## III. THE RATE OF CONVERGENCE OF VARIATIONAL ALGORITHMS

It is of interest to determine the rate of convergence of the variational quantum imaginary evolution [? ] the variational UCC bosonic ansätze, as it determines the number of measurements one needs to perform for the parameter updating until reaching the optimal parameters $\theta^*$. In particular, having an exponential rate of convergence would give an optimal scaling of both the subspace and state fidelity seen in Section IV.E. Analyzing the rate of convergence for heuristic optimization algorithms is currently an active research front [? ? ? ] and thus beyond the scope of this paper. We here give a only heuristic remark on convergence rate using quantum neural tangent kernel (QNTK).

The general strategy for the ground state searching is by updating the parameters as

$$\theta_\mu(t+1) = \theta_\mu(t) - \sum_\nu \eta_\mu(t) A_{\mu\nu}^{-1}(\theta(t)) \frac{\partial}{\partial \theta_\nu} \langle H \rangle_{\theta(t)}, \tag{III.1}$$

where $\theta_\mu(t)$ represents the optimization dynamics with step $t$, the learning rate is given by $\eta_\mu(t)$, and $\langle H \rangle_\theta \equiv \langle \psi(\theta) | H | \psi(\theta) \rangle$. Here, we use $A(\theta(t))$ to represent the metric matrix at the parameter $\theta(t)$. The metric matrix in the gradient descent algorithm is simply the identity matrix. In the following section, we will show this explicit form during the optimization. One can ask if there exists a regime where there is a convergence guarantee and, if so, the rate of convergence for these variational paramterization. One can study this question from over-paramterization using quantum neural tangent kernel (QNTK) [? ]. Further taking $\langle H \rangle_{\theta(t)} \equiv z(\theta(t))$, Eq.(III.1) implies:

$$z(\theta(t+1)) - z(\theta(t)) \equiv \delta z = \sum \frac{\partial z}{\partial \theta_\mu} \delta \theta$$

$$= -\sum_\nu A_{\mu\nu}^{-1}(\theta(t)) \eta_\nu(t) \sum_\mu \frac{\partial z(\theta(t))}{\partial \theta_\mu} \frac{\partial z(\theta(t))}{\partial \theta_\nu} \tag{III.2}$$

Assuming $A$ is the identity matrix and $\eta$ to be parameterization-independent, the resultant is precisely the QNTK defined in [? ]. Note that we can interpret $A_{\mu\nu}^{-1}$ as the *learning rate tensor* as part of the definition of NTK in classical neural networks [? ]. In particular, for the circuits that form at least approximate 2-design that satisfies certain concentration conditions (See in [? ]), the average convergence is of the form

$$\epsilon(\theta(t)) \approx e^{-\gamma t} \epsilon(\theta(0)), \tag{III.3}$$

where $\epsilon \equiv z(\theta) - E_0$, $E_0$ is the ground state energy and

$$\gamma \approx \frac{\eta L \operatorname{tr}(H^2)}{\dim(\mathcal{H})^2}, \tag{III.4}$$

With $L$ being the total count of variational parameters and $H$ be full Hamiltonian. The dimension of the Hilbert space in our case is $n_{\text{cut}}^N$. The exponential convergence rate is guarantee on average in the *over-parametrization regime* where $L \approx \dim(\mathcal{H})^2 / \operatorname{tr}(H^2)$. When $A$ fails to be an identity matrix such as in the case of quantum imaginary time evolution used in the following, no precise analytical convergence guarantee is known. However, it seems to be reasonable to extrapolate the hypothesis that such methods, due to its more physical, geometric nature, would have convergence rates lowered-bounded by the naive gradient descent methods.

Clarifying the gradient descent dynamics would enable us to determine the *query complexity* of the quantum circuits during the optimization phase; that is the number of measurements the variational quantum circuits needed to perform or simply the number of parameter updates. Let us consider the $k$-particle subspace fidelity:

$$f^{(k)}(\tau) = \langle \psi(\theta(\tau)) | \Lambda^{(k)} | \psi(\theta(\tau)) \rangle, \tag{III.5}$$

where the $N$-dimensional projection operator:

$$\Lambda^{(k)} = \sum_i^{D_{\Lambda^{(k)}}} |n_i\rangle \langle n_i|_{\text{k-particle}}, \tag{III.6}$$

where $D_{\Lambda^{(k)}}$ is bounded by $\binom{k+N-1}{k}$. As an intuitive example, we consider the projection to $\Lambda^{(0)} = |\Omega\rangle \langle \Omega|$ and $f^{(0)}(\tau)$. Suppose we initially at $f^{(0)}(\tau_i)$ and we are aiming at fidelity $f^{(0)}(\tau_f) = 1 - \delta$, an important question to ask is what is the query complexity $(\tau_f - \tau_i)/\Delta\tau$? We wish this value to be bounded by $\operatorname{Poly}(N \log n_{\text{cut}})$. Recall from quantum field theories we know that the overlap between physical vacuum state and the free vacuum state $\langle \Omega | 0 \rangle$ is non-vanishing even in the continuum limit. Therefore, having initializing the UCC ansätze $|\psi(\theta(\tau_i))\rangle = c_0 |0\rangle + |\perp\rangle$ for some constant $c_0$ would ensure that the variational algorithms to converge to the physical vacuum state $|\Omega\rangle$. Note that the physical vacuum state is unique so we expect that we can approximate $|\Omega\rangle$ with fidelity $f^{(0)}(\tau_f) = 1 - \delta$ if $\epsilon(\theta(\tau_f)) \leq \delta$ where we can further write from Eq. (III.3):

$$\tau_f - \tau_i \geq \frac{1}{\gamma} \log \frac{\epsilon_0}{\delta} \tag{III.7}$$

for some error $\delta$ and initial error $\epsilon_0$. The log of inverse error looks promising but unfortunately as we computed $1/\gamma$ scales as the dimension of Hilbert space so not super ideal. If $1/\gamma$ were to be bounded by $\operatorname{Poly}(N \log n_{\text{cut}})$, it then follows that the query complexity is polynomially bounded.

To obtain the scaling of $\gamma$ we first assume that the variational ansätze generated by $\hat{T}_1$ and $\hat{T}_2$ (Eq.(25) & (26) in Main text) form an at least approximate 2-design and satisfies the concentration condition [**?** ] in the large $n_{\text{cut}}$ and $N$ limit. Then For each $\hat{T}_\ell$, we have $O(N^\ell n_{\text{cut}}^\ell)$ variational parameters, corresponding to each Pauli operator. As a result, the rate of convergence can be computed:

$$\begin{aligned} \gamma_\ell &\approx O(\frac{\eta N^{3+\ell} n_{\text{cut}}^{\ell+1}}{n_{\text{cut}}^{2N}}) = O(\frac{\eta N^{3+\ell} n_{\text{cut}}^{\ell+1}}{n_{\text{cut}}^{2N}}) \\ &\approx O(\frac{\eta N^{3+\ell}}{n_{\text{cut}}^{(2N-\ell-1)}}). \end{aligned} \tag{III.8}$$

Thus the over-parameterization regime occurs at:

$$\log k + \log \eta + (3+\ell) \log N - (2N - \ell - 1) \log n_{\text{cut}} = 0, \tag{III.9}$$

where the trace norm of the Hamiltonian is bounded by its number count of Pauli operators $O(N^3 n_{\text{cut}})$ and $k$ some constant. Note that this regime fails to hold at large $N$; so the QNTK could only provide some analytic guarantee of exponential convergence rate at some small momentum modes where $N$ scales linearly with the fixed $\ell$. In this regime, we set the learning rate $\log k\eta \approx \ell \frac{\log n_{\text{cut}}}{\log \ell}$. As a result, in general, it is difficult from QNTK to analyze the rate of convergence in the ansätze for a fixed and lower $\ell$ as the constant $k\eta$ has to be linearly increasing with respect to $n_{\text{cut}}$. For NISQ applications, where the number of qubits is within tens of qubits, however, it might be possible to achieve an exponential rate of convergence. For instance, if we set the constant $k = 10000$ and the learning rate $\eta = 0.01$ with $N = 4$, then we have roughly $n_{\text{cut}} = 32$. Thus $\hat{T}_4$ can achieve an exponential rate of convergence, provided the aforementioned assumptions, up to the 20-qubit system.

It should be noted that we expect that NISQ era to have roughly between a few tens to few hundred qubits, clearly above the regime where the rudimentary QNTK theory can provide a theoretical guarantee on logarithms or polynomial query complexity. To this end, to obtain any polynomial query complexity in the variational setting is, to our best knowledge, difficult and largely unexplored; as a result, we leave it to as an open remark. The above analysis could also be understood as a motivation for the significance on researching the direction of QNTK or similar methods, where it could enable us to obtain a rigorous algorithmic prediction instead of relying on heuristic arguments.

## IV. $\phi^4$ THEORY AS AN EFFECTIVE FIELD THEORY

In this appendix, we simply review how the effective description of the classical Ising model in the long distance could be given by field theories, for the convenience of the quantum information science community. The observation where field theory could serve as an effective description of many-body physics has an important significance in theoretical physics, sharpening the importance of simulating quantum field theories.

The $\phi^4$ field is one of the simplest interacting field theories related to the emergent behavior in the physical systems. In this section, we will review that the $d$-dimensional classical Ising model can be described by the $\phi^4$ field at the long distance, following the discussions in [**?** ] rather closely.

The classical magnetism can be described by the Hamiltonian

$$H = \sum_{ij} J_{ij} S_i S_j - \sum_i h_i S_i, \tag{IV.1}$$

where $J_{ij}$ is a translationally invariant correlation matrix describing the interaction of the spins. The classical partition function can be written as

$$\mathcal{Z} = \sum_{\{S_i\}} e^{\sum_{ij} \tilde{J}_{ij} S_i S_j + \sum_i \tilde{h}_i S_i}, \tag{IV.2}$$

where we define $\tilde{J} \equiv -\beta J$ and $\tilde{h}_i \equiv \beta h_i$.

By introducing a continuous auxiliary field $\psi_i$, the partition function is given by

$$\mathcal{Z} = \mathcal{N} \int D\psi \sum_{\{S_i\}} e^{-\frac{1}{4} \sum_{ij} \psi_i [\tilde{J}^{-1}]_{ij} \psi_j + \sum_i S_i(\psi_i + h_i)}, \tag{IV.3}$$

where the Gaussian integral has been evalued as

$$\int Dv(x) \times$$
$$\exp\left[-\frac{1}{2} \int dx dx' v(x) A(x, x') v(x') + \int dx j(x) v(x)\right]$$
$$\propto (\det A)^{-1/2} \exp\left[\frac{1}{2} \int dx dx' j(x) A^{-1}(x, x') j(x')\right].$$

The equation above is under the saddle point approximation, which we preserves the second order of the functional

$$S[x] = S[\bar{x} + y] = S[\bar{x}] + \frac{1}{2} \int dt \int dt' y(t') A(t, t') y(t), \tag{IV.4}$$

The partition function can now be written as [**?** ]

$$\mathcal{Z} = \mathcal{N} \int D\psi \sum_{\{S_i\}} e^{-\frac{1}{2} \sum_{ij} \psi_i [\tilde{J}^{-1}]_{ij} \psi_j + \sum_i S_i(\sqrt{2}\psi_i + h_i)}. \tag{IV.5}$$

Note that the interaction between spins is decoupled by the continuous auxiliary field. In this case, we could work with the multi-dimensional integral $\int D\psi$ instead of the summation over discretized spins

$$\mathcal{Z} = \mathcal{N} \int D\psi e^{-\frac{1}{2} \sum_{ij} (\psi_i - h_i) \tilde{J}_{ij}^{-1} (\psi_j - h_j) + \sum_i \ln(\cosh \sqrt{2}\psi_i)}, \tag{IV.6}$$

where we have used $S_i = \pm 1$, and the normalization factor has absorbed the trivial prefactor. Notice that

$$\ln \cosh(x) = \frac{1}{2}x^2 - \frac{1}{12}x^4 + \cdots.$$

So the Ising model could be described by the $\phi_4$ theory at the long distance.

To make the fourth-order term more clear, we change the integration variable $\psi$ to the field variable via

$$\phi_i \equiv \frac{1}{\sqrt{2}} \sum_j \tilde{J}_{ij}^{-1} \psi_j, \tag{IV.7}$$

and we have

$$\mathcal{Z} = \mathcal{N} \int D\phi e^{-\sum_{ij} \phi_i \tilde{J}_{ij} \phi_j + \sum_i \phi_i h_i + \sum_i \ln \cosh\left(2 \sum_j \tilde{J}_{ij} \phi_j\right)}. \tag{IV.8}$$

We note that the amplitude of the field $\phi$ indicates a measure of magnetism. If the coupling is transnational invariant, which indicates a spatially smooth field, we can capture this feature by the Fourier transformation:

$$\phi_i = \frac{1}{\sqrt{N}} \sum_{\mathbf{k}} e^{-i\mathbf{k} \cdot \mathbf{r}_i} \phi(\mathbf{k}). \tag{IV.9}$$

Given the expansion

$$(\tilde{J}\phi)(\mathbf{k}) = \tilde{J}(\mathbf{k})\phi(\mathbf{k}) = \tilde{J}_0 \phi(\mathbf{k}) + \frac{1}{2}\mathbf{k}^2 \tilde{J}_0'' \phi(\mathbf{k}) + \mathcal{O}\left(\mathbf{k}^4\right), \tag{IV.10}$$

the effective action in the momentum space up to the fourth order is given by

$$\begin{aligned} S[\phi] = \sum_{\mathbf{k}} & \left[\phi_{\mathbf{k}} \left(c_1 + c_2 \mathbf{k}^2\right) \phi_{-\mathbf{k}} + c_3 \phi_{\mathbf{k}} h_{-\mathbf{k}}\right] \\ & + c_4 \sum_{\mathbf{k}_1, \dots, \mathbf{k}_4} \phi_{\mathbf{k}_1} \phi_{\mathbf{k}_2} \phi_{\mathbf{k}_3} \phi_{-\mathbf{k}_1 - \mathbf{k}_2 - \mathbf{k}_3} \left(\mathbf{k}^4, h^2, \phi^6\right), \end{aligned} \tag{IV.11}$$

where

$$\begin{aligned} c_1 &= \tilde{J}_0(1 - 2\tilde{J}_0), \\ c_2 &= \frac{1}{2}\tilde{J}_0''(1 - 4\tilde{J}_0), \\ c_3 &= 1, \end{aligned} \tag{IV.12}$$

which can be determined from the interaction coupling. We can thus see its correspondence to (??), the $\phi^4$ field in the momentum space.

In the real space, the effective action admits the form of $\phi^4$ field as

$$S[\phi] = \int d^d x c_1 \phi^2 + c_2 (\partial \phi)^2 + c_3 \phi h + c_4 \phi^4. \tag{IV.13}$$

When the coupled external field is zero, i.e., $h = 0$, it becomes the prototypical $\phi^4$ field discussed in the main text. Suppose $c_2 > 0$, for $c_1 > 0$, the action has a global minimum at $\phi = 0$, and thus the ground state is in the the paramagnetic phase. While for $c_1 < 0$, the action has two degenerate minima. Thus the phase is spontaneously broken with a ferromagnetic ordering. If the interaction strength $J_{ij} < 0$, the constant $c_1$ is positive at sufficiently low temperature, and it becomes zero at the transition temperature. Therefore, the $\phi^4$ field could capture the critical behavior of the paramagnetic phase to the ferromagnetic phase.


}

\end{document}